\documentclass[letterpaper,10pt]{article}

\title{\LARGE \bf
A Unified Toll Lane Framework for Autonomous and High-Occupancy Vehicles in Interactive Mixed Autonomy}

\usepackage{geometry}
\usepackage{tabu}
\usepackage{epsfig} 
\usepackage{amssymb}  
\usepackage{amsmath,bm}
\let\labelindent\relax
\usepackage{enumitem}
\usepackage{pifont}
\usepackage[dvipsnames]{xcolor}
\usepackage{tikz}
\usepackage{pgfplots}
\usepackage{algorithm2e}
\usepackage{varwidth}
\usepackage{soul}
\usepackage{verbatim}
\usepackage{cite}
\usepackage{booktabs}
\usetikzlibrary{arrows,shapes}
\usepackage{float}

\usepackage{graphics} 
\usepackage{amsmath} 
\usepackage{amssymb}  

\usepackage{amsthm}
\usepackage[scientific-notation=true]{siunitx}
\usepackage[colorlinks]{hyperref}
\usepackage{amsfonts}
\usepackage{tikz}
\usetikzlibrary{arrows.meta}
\usepackage{dsfont}

\usepackage{graphicx}
\usepackage{caption}
\usepackage{subcaption}
\usepackage{multirow}
\usepackage{setspace} 
\usepackage{nomencl}

\newtheorem{theorem}{Theorem}

\newtheorem{proposition}{Proposition}
\theoremstyle{definition}
\newtheorem{example}{Example}
\newtheorem{definition}{Definition}

\theoremstyle{remark}
\newtheorem{remark}{Remark}

\usepackage{placeins}

\definecolor{darkblue}{RGB}{0,101,204}
\definecolor{carorange}{RGB}{255,131,0}

\newcounter{tmp}

\author{Ruolin Li\footnote{R. Li is currently with the Center for Automotive Research at Stanford, Stanford University, and will be with the Department of Civil and Environmental Engineering, University of Southern California starting from August 2024. {\tt\small ruolinli@stanford.edu}} \!, Philip N. Brown\footnote{P. N. Brown is with the Department of Computer Science, University of Colorado Colorado Springs.{\tt\small philip.brown@uccs.edu}} \! and Roberto Horowitz\footnote{R. Horowitz is with the Department of Mechanical Engineering, University of California, Berkeley.{\tt\small horowitz@me.berkeley.edu}}
}

\begin{document}

\maketitle


\thispagestyle{empty}
\pagestyle{plain}

\makenomenclature

\begin{abstract}
In this study, we propose a toll lane framework in the mixture of self-interest-driven autonomous and
high-occupancy vehicles. We examine a scenario where human-driven and autonomous vehicles, categorized by low or high commuter occupancy, share a freeway segment. Autonomous vehicles, capable of maintaining shorter headways, enhance traffic throughput. We introduce a toll lane framework where one lane is designated as restricted and tolled, allowing autonomous vehicles with high occupancy—the category with the highest potential to boost social mobility—to use it without charge, while other vehicles must pay a toll for access. 
Assuming all vehicles are motivated by the goal of minimizing their travel costs, we delve into the resulting lane choice equilibria and uncover their advantageous properties. This is achieved by comparing and ranking various vehicle classes based on their potential to enhance mobility—a concept introduced as the mobility degree.
Utilizing numerical examples, we elucidate the framework's applicability in various design challenges, such as optimal toll setting, occupancy threshold determination, and lane policy design, demonstrating its potential to harmonize high-occupancy and autonomous vehicle integration. 
We further develop an algorithm for assigning rational, differentiated tolls across vehicle classes to efficiently decrease the total commuter delay and assess the impact of potential vehicle misbehavior, specifically unauthorized toll lane usage at reduced toll rates. Our results reveal that self-interest-driven vehicle behavior paradoxically serves as a buffer against the impacts of moderate toll non-compliance, thereby underscoring the framework's inherent resilience.
This investigation stands as the initial systematic exploration of a toll lane framework designed to harmonize the coexistence of autonomous and high-occupancy vehicles within public road systems. It provides valuable perspectives for enhancing traffic management strategies and facilitating the seamless integration of autonomous vehicles into the current transportation infrastructure.
\end{abstract}

\setlength{\parindent}{25pt}

\section{Introduction}\label{intro}

The evolution of autonomous driving technology has spurred significant research and policy development interest in its integration into intelligent transportation systems. Compared to human-driven vehicles, autonomous vehicles offer enhanced reliability by mitigating human operation errors~\cite{farmer2008crash,bagloee2016autonomous} and also contribute to sustainable development goals through the optimization of fuel usage~\cite{asadi2010predictive,luo2010model}. Furthermore, existing studies indicate that autonomous vehicles can boost lane capacity and overall traffic throughput by maintaining shorter headways and forming platoons, compared to their human-driven counterparts~\cite{zohdy2012intersection,talebpour2016influence,lioris2017platoons}.

The efficacy of connected and autonomous vehicles hinges on strategic roadway organization. Specifically, consolidating autonomous vehicles on roadways can enhance platooning efficiency and safety by minimizing interference from human-driven vehicles. Consequently, the implementation of lane policies tailored for autonomous vehicles is crucial, potentially dictating the overall effectiveness of their deployment.  Currently, lane policies for autonomous vehicles fall into two principal categories. The first category is the integrated lane policy~\cite{mehr2019will}. The integrated lane policy indicates that autonomous vehicles travel along with human-driven vehicles in the same group of lanes. Such policies offer convenience but potentially diminish the platooning efficacy and safety of autonomous vehicles. The second category of policies is dedicated lane policies~\cite{mahmassani201650th,ye2018impact,ivanchev2019macroscopic}. Under such policies, some lanes are reserved exclusively for autonomous vehicles, bolstering safety and facilitating vehicle management. However, these policies may have counterproductive implications for social mobility, particularly when the adoption of autonomous vehicles is in its nascent stages, i.e., the penetration rate of autonomous vehicles is low~\cite{doi:10.3141/2622-01,zhong2018assessing}. 
Further, in ~\cite{liu2019strategic}, autonomous vehicle toll lanes are examined, which provide unrestricted access to autonomous vehicles while offering entry to human-driven vehicles for a fee. Such approaches aim to optimize lane usage and alleviate congestion, even when the prevalence of autonomous vehicles is limited.

As the prevalence of autonomous vehicles grows and the need for dedicated lanes becomes more pronounced in terms of safety and mobility benefits, the challenges associated with constructing new dedicated lanes, namely the high cost and extensive time, come to the forefront.
In light of these challenges, recent research has explored the feasibility of repurposing existing lanes, such as those designated for high-occupancy vehicles, for the exclusive use of autonomous vehicles. For example, in~\cite{xiao2019traffic,guo2020leveraging}, simulations and experiments are conducted to investigate the benefit of converting an existing high-occupancy vehicle lane to a dedicated lane for autonomous vehicles. However, such conversion would result in a loss of the benefits brought by a dedicated high-occupancy vehicle lane, especially when a considerable proportion of commuters choose to carpool.


In this study, we aim to synergize the benefits of both high-occupancy vehicle lanes and autonomous vehicle lanes. We examine a highway scenario where four categories of vehicles coexist: low-occupancy human-driven vehicles (HV,LO), high-occupancy human-driven vehicles (HV,HO), low-occupancy autonomous vehicles (AV,LO), and high-occupancy autonomous vehicles (AV,HO). Autonomous vehicles have the capability to enhance traffic throughput by maintaining a shorter headway than their human-driven counterparts. Meanwhile, high-occupancy vehicles carry multiple passengers per vehicle to bypass congestion. We introduce a toll lane framework that allocates a lane exclusively for AV,HOs, i.e., the most capable type of vehicles, at no charge, while providing access to the other three vehicle types for a toll. This model assumes that all vehicles are rational and aim to minimize their travel costs, a composite of their travel time and any tolls paid.

We characterize the common non-unique nature of vehicles' lane choice equilibria and their implications for social mobility enhancement via high-occupancy versus autonomous features. Through numerical examples, we elucidate the toll lane framework's various application venues, including optimal toll setting, occupancy threshold calibration, and policy formulation. We further propose an algorithm that assigns differentiated tolls to various vehicle classes to minimize overall commuter delays and evaluate the system's resilience to potential misbehavior, such as underpayment of tolls. Remarkably, we discover that, under certain conditions, the system's delays are resistant to such misbehavior.
To our knowledge, this is the first comprehensive analysis of a toll lane policy that integrates both autonomous and high-occupancy vehicles, thereby paving the way for more efficient and cooperative use of road infrastructure in the early stage adoption of autonomous vehicles.

As a summary, the contributions of this work can be highlighted as
\begin{itemize}
    \item Introduction of a novel toll lane framework that integrates both autonomous and high-occupancy vehicles on freeways, aiming to enhance traffic throughput by leveraging the shorter headways of autonomous vehicles and promoting higher occupancy.
The framework introduces a concept, the mobility degree to compare various vehicle classes based on their potential to enhance mobility and provides foundational insights for transforming existing road infrastructures to support the integration of autonomous vehicles.
 
    \item Comprehensive investigation of the lane choice behavior under the proposed toll lane framework, identifying equilibria and their desirable properties to enable efficient optimization techniques and versatile applications, such as toll design, occupancy threshold determination, and policy design exploration.

\item Proposal of an algorithm to calculate differentiated tolls for various vehicle classes in order to efficiently decrease total commuter delay, addressing the challenge of vehicle misbehavior and unauthorized toll lane usage, offering insights into the resilience of the proposed framework against toll violations.

\item Revelation of selfish behavior's dual role in interactive mixed autonomy. Selfish behavior, traditionally viewed as detrimental to traffic efficiency, can under certain conditions serve as a beneficial buffer. This complements conventional perceptions and suggests the complexity of interactions between human-driven and autonomous vehicles.
\end{itemize}

Building on the groundwork established in our previous publication~\cite{9564777}, this paper introduces a significant conceptual advancement with the introduction of the {\bf mobility degree}. This novel concept allows for a more nuanced understanding of vehicle characteristics within mixed autonomy traffic systems. We have comprehensively reformulated the framework, integrating this concept to provide a more structured foundation for our analysis.
Further, we have expanded our investigation to include an in-depth resilience analysis, meticulously examining how the system withstands and adapts to toll violation disruptions. Central to this exploration is the added insight into the paradoxical role of selfish behavior among vehicles. Contrary to traditional views that label such behavior as purely detrimental, our findings illuminate its capacity to act as a buffer, enhancing system resilience under certain conditions.

This paper is organized as follows. In Section~\ref{sec:model}, we introduce the toll lane framework and vehicles' lane choice model. In Section~\ref{sec:NE}, we examine the properties of the resulting lane choice equilibria. In Section~\ref{sec:design}, we clarify how the toll lane framework can be applied to finding the optimal toll/occupancy threshold/lane policy that minimizes the total commuter delay. In Section~\ref{sec:diff_toll}, we propose an efficient method to decrease the total commuter delay by differentiating the tolls. In Section~\ref{sec:misbehavior}, we examine the potential misbehavior of vehicles and discuss the impact of vehicles' misbehavior on the total delay of all commuters. Finally, in Section~\ref{sec:future}, we conclude this work and discuss future implications.

\section{The Toll Lane Choice Model}\label{sec:model}
In this section, we propose a toll lane framework designed to accommodate vehicles with distinct attributes of autonomy and occupancy. This framework aims to analyze and predict the lane choice behavior of such vehicles at equilibrium. To this end, we employ a model reminiscent of Wardrop Equilibrium~\cite{wardrop1952some}, allowing us to characterize these behaviors under steady-state conditions with precision and clarity.

\begin{figure}
\centering
 \includegraphics[width=0.65\textwidth]{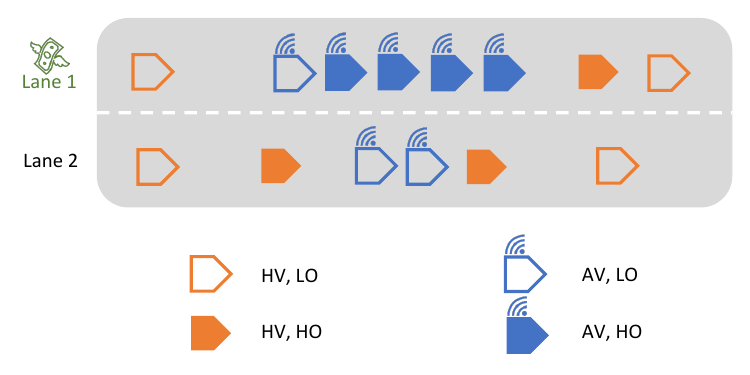}
\caption{Overview of the Problem Setting. This diagram illustrates the lane usage strategy within the proposed framework, where Autonomous Vehicles with High Occupancy (AV,HO) are granted toll-free access to Lane 1. In contrast, vehicles belonging to the other three classes are presented with a choice: to pay a toll for the privilege of using Lane 1 or to navigate Lane 2 without any toll requirement.}
\label{fig:diverge_basic}
\end{figure}

In the context of the highway segment illustrated in Figure~\ref{fig:diverge_basic}, let $\mathcal{I}=\{1,2\}$ denote the set of lane indices. Specifically, Lane 1 is designated as the reserved toll lane(s), offering exclusive, toll-free access to certain classes of vehicles, while permitting access to others upon payment of a toll. Lane 2 represents a regular lane(s), accessible to vehicles of all classes without any toll requirement.

We consider that four classes of vehicles are sharing the roads: human-driven vehicles with low occupancy (HV,LO), human-driven vehicles with high occupancy (HV,HO), autonomous vehicles with low occupancy (AV,LO) and autonomous vehicles with high occupancy (AV,HO). We denote 
\begin{align}
    \mathcal{P}:= \{\rm HV,LO;HV,HO;AV,LO; AV,HO\}
\end{align}
to be the set of all classes of vehicles. 

\begin{remark}[High-occupancy versus low-occupancy vehicles]
High-occupancy vehicles have at least two commuters per vehicle, whereas low-occupancy vehicles have at most one commuter per vehicle. Let $n^{\text{HO}}$ be the average number of commuters carried by a high-occupancy vehicle, and $n^{\text{LO}}$ be the average number of commuters carried by a low-occupancy vehicle. We have $n^{\text{HO}} \geq 2$, $n^{\text{LO}} > 0$ and $ n^{\text{HO}}> n^{\text{LO}}$.
\end{remark}

\begin{remark}[Human-driven versus autonomous vehicles]
We assume in this paper that autonomous vehicles maintain a shorter headway than human-driven vehicles (see experimental evidence in~\cite{smith2020improving}). Let the ratio between the average headway of autonomous vehicles and the average headway of human-driven vehicles be $\mu$, we have $\mu \in (0,1)$. Noticeably, $\mu$ coincides with the concept, capacity symmetry degree introduced and studied in~\cite{mehr2019will} and~\cite{li2020game}.
\end{remark}

Throughout the work, we consider a daily commuting situation, where the commuter demands on the segment of highway can be considered inelastic for a certain period. 
We denote the {\bf commuter demands} as
\begin{align}
    \mathbf{d} := \left(d^p, p\in \mathcal{P} \right), 
\end{align}
where $d^{p}\ge 0$ represents the demand of commuters who either drive or commute in a vehicle of class $p\in \mathcal{P}$. Consequently, the {\bf vehicle demands} can be written as
\begin{align}
\label{eq:vd}
\mathbf{d}_{\text{v}} := \left(d_\text{v}^p, p\in \mathcal{P} \right) :=
\left( \frac{d^{\rm HV,LO}}{n^{\text{LO}}},\, \frac{d^{\rm HV,HO}}{n^{\text{HO}}}, \,
     \frac{d^{\rm AV,LO}}{n^{\text{LO}}},\, \frac{d^{\rm AV,HO}}{n^{\text{HO}}} \right). 
\end{align}

In this paper, we use the commonly-received volume-capacity delay models, such as BPR functions~\cite{roads1964traffic}, in which the travel delay is a continuous and increasing function of the flow-capacity ratio. Thus, in terms of their impact on delay, a certain flow of autonomous vehicles with a shorter headway can be seen as a smaller effective flow of human-driven vehicles (check the similar concept in~\cite{liu2019strategic}). The shorter autonomous vehicles' headway is, the smaller the effective flow is. We then define $\delta^p$ to be the effective demand for vehicles of class $p\in \mathcal{P}$ and collect {\bf effective vehicle demands} in the vector: 
\begin{align}
\label{eq:effdem}
\boldsymbol \delta := \left(\delta^p, p\in \mathcal{P} \right):= \left( \frac{d^{\rm HV,LO}}{n^{\text{LO}}},\, \frac{d^{\rm HV,HO}}{n^{\text{HO}}}, \,
     \mu \frac{d^{\rm AV,LO}}{n^{\text{LO}}},\, \mu \frac{d^{\rm AV,HO}}{n^{\text{HO}}} \right). 
\end{align}
Notice that effective vehicle demand $\delta^p$ describes the substantive impact of the demand of class $p$ vehicles on the traffic delay at the segment of the highway. Furthermore, we are ready to present the key definition in the work.

\begin{definition} [\bf Mobility Degree] \label{def:md}
    Mobility Degree $\nu^p$ of a certain class of vehicles $p\in \mathcal{P}$ is defined as the ratio between its commuter demand and its effective vehicle demand, i.e., 
    \begin{align}
        \nu^p := \frac{d^p}{\delta^p},\ p\in \mathcal{P}.
    \end{align}
\end{definition}

\begin{remark}[Mobility Degree]
    Mobility Degree of a class of vehicles indicates its competency in enhancing mobility by its attributes in occupancy and autonomy. A larger mobility degree indicates a smaller effective vehicle demand under a certain commuter demand, implies a smaller contribution to the delay, and represents a greater competency in increasing social mobility.
\end{remark}

For all classes of vehicles in $\mathcal{P}$, their Mobility Degrees are calculated as the following:
\begin{align}
\nu^{\rm HV,LO} &= n^{\text{LO}},\\
\nu^{\rm HV,HO} &= n^{\text{HO}},\\
\nu^{\rm AV,LO} &= \frac{n^{\text{LO}}}{\mu},\\
\nu^{\rm AV,HO} &= \frac{n^{\text{HO}}}{\mu}.
\end{align}
We can further rank different classes of vehicles' Mobility Degree as:
\begin{align}\label{eq:def_nu}
\nu^{\rm HV,LO}  <
    \left\{\nu^{\rm HV,HO},
    \ \nu^{\rm AV,LO} \right\}<
    \nu^{\rm AV,HO} .
\end{align}
Notice that we do not assume whether the mobility degree of HV,HO vehicles is larger than that of AV,LO, or vice-versa. Both cases will be considered in the following parts.

Among the four classes of vehicles categorized within the framework of our study, AV,HOs have the highest Mobility Degree (as ranked in~\eqref{eq:def_nu}). This distinction arises from AV,HOs' dual capacity to transport multiple commuters per vehicle and to maintain a shorter headway than human-driven vehicles. These capabilities uniquely position AV,HOs to significantly enhance the overall mobility of the freeway segment. To leverage this potential, our model permits AV,HOs to utilize Lane 1 without incurring toll charges. Recognizing that an increased presence of autonomous vehicles not only elevates safety levels but also facilitates the formation of vehicle platoons, {\bf we route all AV,HOs on Lane 1 without charging a toll.} The remaining three classes of vehicles face a choice: they can opt to pay a toll for access to Lane 1 or navigate freely on Lane 2. In the following discussion, we will elucidate how their choices influence vehicle distribution and corresponding mobility outcomes. We gather the three classes of {\bf decision-making vehicles} in the following set:
\begin{eqnarray}
\label{eq:Pbar}
     \Bar{\mathcal{P}}:=\{\rm HV,LO;HV,HO;AV,LO\} \subset \mathcal{P}.
\end{eqnarray}
Notice that we exclude AV,HOs from this set since we assume that they will always travel on Lane 1.

Given lane $i\in \mathcal{I}$, we denote the flow of vehicles of class $p\in \Bar{\mathcal{P}}$ on that lane as $f_i^p$, and the flows of decision-making vehicles on lane $i\in \mathcal{I}$ as $\mathbf{f}_i$ characterized as below:
\begin{align}
    \mathbf{f}_1 &:= \left(f_1^{\rm HV,LO},\ f_1^{\rm HV,HO},\ f_1^{\rm AV,LO}\right),\label{eq:fdef}\\
    \mathbf{f}_2 &:= \left(f_2^{\rm HV,LO},\ f_2^{\rm HV,HO},\ f_2^{\rm AV,LO}\right).\nonumber
\end{align}
Together, the distribution of decision-making vehicles is represented by vector $\mathbf{f}$ characterized as:
\begin{align}\label{eq:ffffff}
    \mathbf{f} &:= \left(\mathbf{f}_1, \mathbf{f}_2\right)\in \mathbb{R}_+^{6}.
\end{align}
Notice that the flows in $\mathbf{f}$ must satisfy:
\begin{align}
    &\sum_{i\in \mathcal{I}}f_i^{\rm HV,LO} = d_\text{v}^{\rm HV,LO},\nonumber\\
    &\sum_{i\in \mathcal{I}}f_i^{\rm HV,HO} =d_\text{v}^{\rm HV,HO},\label{eq:2}\\
    &\sum_{i\in \mathcal{I}}f_i^{\rm AV,LO} =d_\text{v}^{\rm AV,LO},\nonumber \\
    \text{and}\ \ f_i^{\rm HV,LO}\geq 0,\ &f_i^{\rm HV,HO}\geq 0,\ f_i^{\rm AV,LO}\geq 0,\ \forall i\in \mathcal{I},\label{eq:88}
\end{align}
where $d_\text{v}^p$ are vehicle demands of vehicles class $p$ defined in~\eqref{eq:vd}.
Similar to the concept of effective vehicle demands in \eqref{eq:effdem}, we use {\bf effective vehicle flow} $\phi_i$ in each lane $i\in \mathcal{I}$ to account for the substantive impact of mixed autonomy vehicles on the travel delay in each lane, and we have
\begin{align}
    \phi_1 &:=f_1^{\rm HV,LO}+f_1^{\rm HV,HO}+\mu f_1^{\rm AV,LO}+\delta^{\rm AV,HO},\label{eq:f1}\nonumber\\
\phi_2 &:=f_2^{\rm HV,LO}+f_2^{\rm HV,HO}+\mu f_2^{\rm AV,LO}.
\end{align}
Naturally, we have
\begin{align}
    \phi_1 + \phi_2 = \sum_{p\in \mathcal{P}}\delta^{p}.\label{eq:sum_phi}
\end{align}
Moreover, the effective flows should fulfill the following constraints:
\begin{eqnarray}\label{eq:range_phi}
     \phi_1 \in \left[\delta^{\rm AV,HO},\sum_{p\in \mathcal{P}}\delta^{p}\right] \:\:{\mbox{and}\:\:
     \phi_2 \in \left[0,\sum_{p\in \mathcal{P}}\delta^{p}-\delta^{\rm AV,HO}\right] }, 
\end{eqnarray} 
where $\delta^p$ are effective vehicle demands of vehicle class $p$ defined in~\eqref{eq:effdem}.

In this work, we assume that all decision-making vehicles behave selfishly by always choosing the lane that minimizes their travel cost. The travel cost takes into account the travel delay and any toll expense that exists. We first assume that a uniform toll price $\tau> 0$ is charged to decision-making vehicles that choose to travel on lane 1, while no toll is charged to vehicles traveling on lane 2. Thus, for decision-making vehicles traveling on lane 1, the travel cost equals the sum of the travel delay on lane 1 and the toll, whereas, for decision-making vehicles traveling on lane 2, the travel cost is equal to the travel delay. For lane $i\in \mathcal{I}$, we denote $C_i$ as the travel cost, and $D_i$ as the travel delay.
Recall that in this paper, we use volume-capacity delay models, such as BPR functions~\cite{roads1964traffic}, in which the travel delay on lane $i$, $D_i$ is a continuous and increasing function of the effective vehicle flow $\phi_i$. We then have
\begin{align}
    C_1(\mathbf{f}) &=D_1(\phi_1)+\tau,\nonumber\\
    C_2(\mathbf{f}) &=D_2(\phi_2).\label{eq:tc11}
\end{align}
More details of the capacity and delay modeling of mixed autonomy can be found in~\cite{lazar2017capacity}.
\begin{remark}[Homogeneous standardized Value of Time of mixed autonomy]
In this study, we assume a homogeneous value of travel time for vehicles in mixed autonomy, standardized as 1. Future adjustments might be necessary if the value of time for Autonomous Vehicles significantly changes. Exploring the full potential and various applications of autonomous vehicles, though beyond the scope of this paper, presents valuable avenues for future research.
\end{remark}

Let the tuple $G=\left(\mathbf{D},\mathbf{d},n^{\text{LO}}, n^{\text{HO}},\mu,\tau\right)$ represent a segment of highway shown in Figure~\ref{fig:diverge_basic} with the delay models $\mathbf{D}:=(D_i,i\in\mathcal{I} = \{1,2\})$, commuter demands $\mathbf{d}$, average occupancy for low-occupancy vehicles $n^{\text{LO}}$, average occupancy for high-occupancy vehicles $n^{\text{HO}}$, a reduced headway ratio $\mu$ for autonomous vehicles, and a uniform toll price $\tau$ charged to decision-making vehicles. The toll lane choice equilibrium of the selfish decision-making vehicles can then be modeled as a Wardrop equilibrium~\cite{wardrop1952some} as shown below.

\begin{definition}[{\bf Toll lane choice equilibrium} of selfish decision-making vehicles]\label{def:wdp_basic}
For a segment of highway $G=\left(\mathbf{D},\mathbf{d},n^{\text{LO}}, n^{\text{HO}},\mu,\tau\right)$, a feasible flow distribution vector $\mathbf{f}$ is a lane choice equilibrium if and only if for any vehicle class $p\in \Bar{\mathcal{P}}$, we have
\begin{subequations}\label{eq:eq_def}
    \begin{align}
    f_1^{p} (C_1(\mathbf{f}) - C_2(\mathbf{f})) &\leq 0 ,\\
    f_2^{p} (C_2(\mathbf{f}) - C_1(\mathbf{f}))&\leq 0.
    \end{align}
\end{subequations}
\end{definition}
\noindent The definition presents sufficient and necessary conditions for the choice equilibrium. Specifically, at the choice equilibrium, if the travel cost of lane 1 is higher than the travel cost of lane 2, then all of the decision-making vehicles would travel on lane 2; if the travel cost of lane 1 is lower than the travel cost of lane 2, then all of the decision-making vehicles would choose to pay the toll and travel on lane 1; if the travel cost of lane 1 is equal to the travel cost of lane 2, then any vehicle of the decision-making vehicles could travel either on lane 1 or lane 2. Reasoning reversely, if any of the decision-making vehicles are on lane 1, then the travel cost of lane 1 cannot be higher than the cost of lane 2; if any of the decision-making vehicles are on lane 2, then the travel cost of lane 2 cannot be higher than the cost of lane 1; if any class of decision-making vehicles uses both lane 1 and lane 2, then the travel cost of lane 1 and lane 2 must be equal.

To conclude this section, we further introduce the total delay of all commuters $J(\mathbf{f})$ as a metric that evaluates the social mobility outcome. The total delay of all commuters $J(\mathbf{f})$ is essentially taking the sum of travel delay of all commuters traveling in vehicles of class $p\in \mathcal{P}$:
\begin{align}\label{eq:social_cost}
   J(\mathbf{f})=
   &\left[n^{\text{LO}} \left(f_1^{\rm HV,LO}+f_1^{\rm AV,LO}\right)+n^{\text{HO}}\left(f_1^{\rm HV,HO}+d_\text{v}^{\rm AV,HO} \right)\right] D_1(\phi_1) \nonumber\\
   &+ \left[n^{\text{LO}} \left(f_2^{\rm HV,LO}+f_2^{\rm AV,LO}\right)+n^{\text{HO}}f_2^{\rm HV,HO}\right]D_2(\phi_2).
\end{align}
A selfish lane choice equilibrium described in Definition~\ref{def:wdp_basic} is usually not socially optimal, i.e, not minimizing the total delay of all commuters. We will examine such properties in the following sections.

\section{Equilibrium properties}\label{sec:NE}
In this section, we investigate the essential properties of the lane choice equilibria outlined in Definition~\ref{def:wdp_basic}. Our analysis begins by confirming their existence, then proceeds to identify conditions that guarantee uniqueness, and finally characterizes the typical non-unique scenarios.

According to the core theorem in~\cite{braess1979existence}, we give the following proposition without proof.
\begin{proposition}[Existence]\label{thm:existence}
For a tolled segment of highway $G=\left(\mathbf{D},\mathbf{d},n^{\text{LO}}, n^{\text{HO}},\mu,\tau\right)$, there always exists at least one lane choice equilibrium as described in Definition~\ref{def:wdp_basic}.
\end{proposition}

The subsequent theorem illustrates that the uniqueness of the resulting lane choice equilibrium occurs under two specific conditions: either when all three classes of decision-making vehicles opt to utilize the same lane at equilibrium, or when there is a single, homogeneous class of decision-making vehicles.

\begin{theorem}[Conditions for uniqueness]\label{thm:uniqueness}
For a tolled segment of highway $G=\left(\mathbf{D},\mathbf{d},n^{\text{LO}}, n^{\text{HO}},\mu,\tau\right)$, the lane choice equilibrium as described in Definition~\ref{def:wdp_basic} is unique if and only if at least one of the following conditions holds:
\begin{description}
    \item[1)] $\tau \geq D_2\left(\sum_{p\in  \mathcal{P}}\delta^{p}-\delta^{\rm AV,HO}\right)-D_1\left(\delta^{\rm AV,HO}\right),$
    \item[2)] $\tau \leq D_2\left(0\right)-D_1\left(\sum_{p\in  \mathcal{P}}\delta^{p}\right).$
    \item[3)] There exists at most one class of decision-making vehicles $p\in \Bar{\mathcal{P}}$, for which the commuter demand is strictly positive, i.e., $d^p>0$.
\end{description}
\end{theorem}

\begin{proof}

\begin{figure*}
\centering
     \begin{subfigure}[b]{0.32\textwidth}
         \centering
         \includegraphics[width=\textwidth,height=150pt]{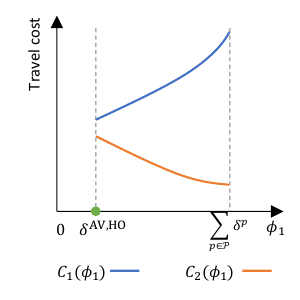}
         \caption{$C_1(\phi_1)\geq C_2(\phi_1)$, $\forall \phi_1\in [\delta^{\rm AV,HO},\sum_{p\in  \mathcal{P}}\delta^{p}]$.}
         \label{fig:case a}
     \end{subfigure}
     \hfill
    \begin{subfigure}[b]{0.32\textwidth}
         \centering
         \includegraphics[width=\textwidth,height=150pt]{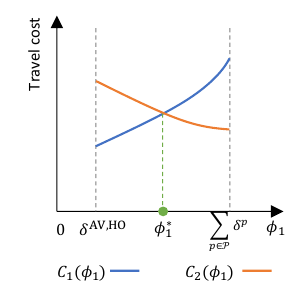}
         \caption{$C_1(\phi_1)$ and $C_2(\phi_1)$ intersect at $\phi_1^*\in \left(\delta^{\rm AV,HO},\sum_{p\in  \mathcal{P}}\delta^{p}\right)$.}
         \label{fig:case b}
     \end{subfigure}
     \hfill
    \begin{subfigure}[b]{0.32\textwidth}
         \centering
         \includegraphics[width=\textwidth,height=150pt]{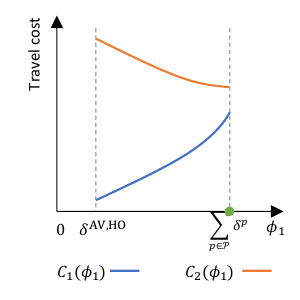}
         \caption{$C_1(\phi_1)\leq C_2(\phi_1)$, $\forall \phi_1\in [\delta^{\rm AV,HO},\sum_{p\in  \mathcal{P}}\delta^{p}]$.}
         \label{fig:case c}
     \end{subfigure}
    \caption{All possible sketches of the travel cost on both lanes. Resulting lane choice equilibria are indicated by the green dots. Non-unique equilibria only exist in case (b).}\label{fig:JJ}
\end{figure*}

The travel cost on Lane 1, $C_1(\mathbf{f})$ is a continuous, increasing function of $\phi_1$, and therefore can be written as $C_1(\phi_1)$. Similarly, $C_2(\mathbf{f})$ is a continuous, increasing function of $\phi_2$. According to Equation~\eqref{eq:sum_phi}, with slight abuse of notation,  we can also treat the travel cost on Lane 2, $C_2(\mathbf{f})$ as a continuous, decreasing function of Lane 1's effective flow $\phi_1$, written as $C_2(\phi_1)$. 
Three possible sketches of $C_1(\phi_1)$ and $C_2(\phi_1)$, for $\phi_1\in [\delta^{\rm AV,HO},\sum_{p\in  \mathcal{P}}\delta^{p}]$ are shown in Figure~\ref{fig:JJ}. In case (a), $C_1(\phi_1)\geq C_2(\phi_1)$ for any possible $\phi_1\in [\delta^{\rm AV,HO},\sum_{p\in  \mathcal{P}}\delta^{p}]$, thus all of the three classes of decision-making vehicles would use Lane 2, and the lane choice equilibrium is unique at $\mathbf{f}_1= \left(0,\ 0,\ 0\right)$. In case (c), $C_1(\phi_1)\leq C_2(\phi_1)$ for any possible $\phi_1\in [\delta^{\rm AV,HO},\sum_{p\in  \mathcal{P}}\delta^{p}]$, thus all of the three classes of decision-making vehicles would use Lane 1, and the lane choice equilibrium is unique at $\mathbf{f}_1=\left(\frac{d^{\rm HV,LO}}{n^{\text{LO}}},\ \frac{d^{\rm HV,HO}}{n^{\text{HO}}},\ \frac{d^{\rm AV,LO}}{n^{\text{LO}}}\right)$. 

In case (b), the resulting lane choice equilibria are in general not unique. Possible lane choice equilibria in case (b) that satisfy Definition~\ref{def:wdp_basic} must satisfy
\begin{align}
    C_1(\phi_1^*)= C_2(\phi_1^*),\label{eq:f1*}
\end{align}
where $\phi_1^*\in\left(\delta^{\rm AV,HO},\sum_{p\in  \mathcal{P}}\delta^{p}\right)$ is the value of $\phi_1$ at the equilibria, which can be readily solved by a golden-section search algorithm~\cite{d6e06843-9f44-3b92-881d-f5d31b55c5e7} given the highway tuple $G$. Thus according to Equation~\eqref{eq:f1}, the resulting equilibria must satisfy
\begin{eqnarray}
\label{eq:Sequilibria}
  f_1^{\rm HV,LO}+f_1^{\rm HV,HO}+\mu f_1^{\rm AV,LO} &=&   \phi_1^*-\delta^{\rm AV,HO}, \nonumber\\
 f_2^{\rm HV,LO}+f_2^{\rm HV,HO}+\mu f_2^{\rm AV,LO} &=&  \sum_{p\in \mathcal{P}}\delta^{p}-\phi_1^*.\label{eq:range}
\end{eqnarray}
All resulting equilibria should also satisfy the feasibility conditions. Therefore, the resulting equilibria lie in a simplex $\mathcal{S}\subset \mathbb{R}^6_+$ which can be characterized as
\begin{align}
    \mathcal{S}:=\{\mathbf{f}\in \mathbb{R}^6_+:\mathbf{f}\ \text{satisfies}~\eqref{eq:2},~\eqref{eq:88}\ \text{and} ~\eqref{eq:range}\}.
\label{eq:S}
\end{align}
The simplex reduces to a single point if and only if there is at most one class of decision-making vehicles $p\in \Bar{\mathcal{P}}$ with a strictly positive commuter demand, i.e., $d^p>0$. 
\end{proof}

\begin{remark}[{\bf Non-unique lane choice equilibria}]
    For a tolled segment of highway $G=\left(\mathbf{D},\mathbf{d},n^{\text{LO}}, n^{\text{HO}},\mu,\tau\right)$, when no condition in Theorem~\ref{thm:uniqueness} is met, the resulting lane choice equilibria are non-unique, and they can be characterized by the simplex $\mathcal{S}$~\eqref{eq:S}. Notice that the simplex $\mathcal{S}$ can be fully characterized by the highway segment tuple $G$. Also notice that, at the equilibria in the simplex $\mathcal{S}$, according to Equation~\eqref{eq:f1*}, where $\tau>0$, the travel delay on Lane 1 is strictly smaller than the travel delay on Lane 2. Therefore, all AV,HOs will naturally accumulate on Lane 1 due to zero toll, even if we do not deliberately route them.
\end{remark}

When there are multiple possible lane choice equilibria, the total commuter delay at the equilibrium can be ambiguous. Therefore, in this paper, we mainly study the multiple equilibria in the simplex $\mathcal{S}$ and in the next theorem we will characterize the multiple equilibria in the simplex $\mathcal{S}$ that respectively yield the largest and the smallest total commuter delay.

For a tolled segment of highway $G=\left(\mathbf{D},\mathbf{d},n^{\text{LO}}, n^{\text{HO}},\mu,\tau\right)$ with non-unique lane choice equilibria characterized by the simplex $\mathcal{S}$~\eqref{eq:S},
let $\mathbf{f}^+\in \mathcal{S}$ be the {\bf worst-case} equilibrium that reaches the {\bf largest} total commuter delay, i.e., 
\begin{align}
    J\left(  \mathbf{f}\right)\leq J\left(  \mathbf{f}^+\right),\ \forall \mathbf{f}\in \mathcal{S},
\end{align}
and let $\mathbf{f}^-\in \mathcal{S}$ be the {\bf best-case} equilibrium that reaches the {\bf smallest} total commuter delay, i.e.,
\begin{align}
    J\left(  \mathbf{f}\right)\geq J\left(  \mathbf{f}^-\right),\ \forall \mathbf{f}\in \mathcal{S},
\end{align}
where $\mathbf{f}$, as defined in Eq.~\eqref{eq:ffffff}, is any equilibrium flow, and the total commuter delay $J\left(  \mathbf{f}\right)$ is given by Eq.~\eqref{eq:social_cost}.

\begin{theorem}[{\bf Characterization of the best-case and the worst-case lane choice equilibrium}]\label{thm:characterization}
For a tolled segment of highway $G=\left(\mathbf{D},\mathbf{d},n^{\text{LO}}, n^{\text{HO}},\mu,\tau\right)$ with non-unique lane choice equilibria characterized by the simplex $\mathcal{S}$~\eqref{eq:S},
\begin{description}
    \item[1)] If $\frac{n^{\text{HO}}}{n^{\text{LO}}}>\frac{1}{\mu}$ (or equivalently, $\nu^{\rm HV,HO} > \nu ^{\rm AV,LO}$),  we have
    \begin{subequations}
    \begin{align}
    \mathbf{f}^+_1=\left(\underset{\mathbf{f}\in \mathcal{S}}{\text{\rm max}}\ f_1^{\rm HV,LO},\ \underset{\mathbf{f}\in \mathcal{S}}{\text{\rm min}}\ f_1^{\rm HV,HO},\ *\right), \label{eq:f1+nuHVgnuav}\\
        \mathbf{f}^-_1=\left(\underset{\mathbf{f}\in \mathcal{S}}{\text{\rm min}}\ f_1^{\rm HV,LO},\ \underset{\mathbf{f}\in \mathcal{S}}{\text{\rm max}}\ f_1^{\rm HV,HO},\ *\right).
    \end{align}
    \end{subequations}
    \item[2)] If $\frac{n^{\text{HO}}}{n^{\text{LO}}}<\frac{1}{\mu}$  (or equivalently, $\nu^{\rm HV,HO} < \nu ^{\rm AV,LO}$), we have
    \begin{subequations}
    \begin{align}
        \mathbf{f}^+_1=\left( \underset{\mathbf{f}\in \mathcal{S}}{\text{\rm max}}\ f_1^{\rm HV,LO},\ *,\ \underset{\mathbf{f}\in \mathcal{S}}{\text{\rm min}}\ f_1^{\rm AV,LO}\right),\\
        \mathbf{f}^-_1=\left( \underset{\mathbf{f}\in \mathcal{S}}{\text{\rm min}}\ f_1^{\rm HV,LO},\ *,\ \underset{\mathbf{f}\in \mathcal{S}}{\text{\rm max}}\ f_1^{\rm AV,LO}\right).
    \end{align}
    \end{subequations}
    \item[3)] If $\frac{n^{\text{HO}}}{n^{\text{LO}}}=\frac{1}{\mu}$  (or equivalently, $\nu^{\rm HV,HO} = \nu ^{\rm AV,LO}$), we have
    \begin{subequations}
    \begin{align}
    \mathbf{f}^+_1=\left(\underset{\mathbf{f}\in \mathcal{S}}{\text{\rm max}}\ f_1^{\rm HV,LO},\ *,\ *\right),\\
        \mathbf{f}^-_1=\left(\underset{\mathbf{f}\in \mathcal{S}}{\text{\rm min}}\ f_1^{\rm HV,LO},\ *,\ *\right).
    \end{align}
    \end{subequations}
\end{description}
\noindent Symbol $*$ in the equations signifies that the quantity can be any value that fulfills feasibility conditions, i.e., $\mathbf{f}^+\in \mathcal{S}$ and $\mathbf{f}^-\in \mathcal{S}$.
\end{theorem}

\begin{proof}
In this proof, we will only provide a detailed derivation for $\mathbf{f}^+$ in Eq. \eqref{eq:f1+nuHVgnuav}, when $\frac{n^{\text{HO}}}{n^{\text{LO}}}>\frac{1}{\mu}$. Similar derivations can be used to prove the other results in the theorem, and are omitted due to clarity considerations.

Assume that $\frac{n^{\text{HO}}}{n^{\text{LO}}}>\frac{1}{\mu}$, and $\mathbf{f^+}=\left( \mathbf{f}^+_1, \mathbf{f}^+_2\right)$, where 
\begin{eqnarray*}
\mathbf{f}^+_1 = \left ( \max_{\mathbf{f}\in \mathcal{S}}  f_1^{\rm HV,LO},\ \underset{\mathbf{f}\in \mathcal{S}}{\text{\rm min}}\ f_1^{\rm HV,HO},\ f_1^{+{\rm AV,LO}} \right) \:\: \mbox{and}\:\:
\mathbf{f}^+_2 =\left(  f_2^{+{\rm HV,LO}},\  f_2^{+{\rm HV,HO}},\ f_2^{+{\rm AV,LO}} \right).
\end{eqnarray*} 
Further let $\mathbf{f}=\left( \mathbf{f}_1, \mathbf{f}_2\right)\in \mathcal{S}$, we aim to prove that $J\left(  \mathbf{f}\right)\leq J\left(  \mathbf{f}^+\right)$ and $\mathbf{f^+}\in \mathcal{S}$ always hold.

From Eq. \eqref{eq:social_cost} we have

\begin{eqnarray}\label{eq:dsocial_cost}
   J(\mathbf{f}) - J(\mathbf{f^+}) &=&
   \left[ n^\text{LO}\left(f_1^{\rm HV,LO}+f_1^{\rm AV,LO}\right)+n^\text{HO} f_1^{\rm HV,HO}  \right] D_1(\phi_1^*)\nonumber\\
   && + \left[n^\text{LO}\left(f_2^{\rm HV,LO}+f_2^{\rm AV,LO}\right)+n^\text{HO}f_2^{\rm HV,HO}\right]D_2(\phi_2^*) \\
   && - \left[ n^\text{LO}\left( \max_{\mathbf{f}\in \mathcal{S}} f_1^{\rm HV,LO}+f_1^{+\rm AV,LO}\right)+n^\text{HO}   \min_{\mathbf{f}\in \mathcal{S}} f_1^{\rm HV,HO}  \right] D_1(\phi_1^*)\nonumber\\
   && - \left[n^\text{LO}\left(f_2^{+\rm HV,LO}+f_2^{+\rm AV,LO}\right)+n^\text{HO}f_2^{+\rm HV,HO}\right]D_2(\phi_2^*). \nonumber
\end{eqnarray}
Since both $\mathbf{f}$ and $\mathbf{f}^+$  satisfy Equation~\eqref{eq:range},  we have
\begin{eqnarray}
f_1^{\rm AV,LO} - f_1^{+\rm AV,LO} &=& \frac{1}{\mu}
\left [ \left (  \max_{\mathbf{f}\in \mathcal{S}} f_1^{\rm HV,LO} - f_1^{\rm HV,LO} \right ) + \left (  \min_{\mathbf{f}\in \mathcal{S}} f_1^{\rm HV,HO} - f_1^{\rm HV,HO} \right ) \right ],\\
f_2^{\rm AV,LO} - f_2^{+\rm AV,LO} &=&\frac{1}{\mu}
\left [ \left (    f_2^{+\rm HV,LO} - f_2^{\rm HV,LO} \right ) + \left (    f_2^{+\rm HV,HO} - f_2^{\rm HV,HO} \right ) \right ].
\end{eqnarray}
Furthermore, from Eqs. \eqref{eq:2}, we have
\begin{eqnarray}
f_2^{\rm HV,LO} - f_2^{+\rm HV,LO} &=& \max_{\mathbf{f}\in \mathcal{S}} f_1^{\rm HV,LO} - f_1^{\rm HV,LO}, \\
f_2^{\rm HV,HO} - f_2^{+\rm HV,HO} &=& \min_{\mathbf{f}\in \mathcal{S}} f_1^{\rm HV,HO} - f_1^{\rm HV,HO}\,.
\end{eqnarray}
Therefore,
\begin{align}
    J\left(  \mathbf{f}\right)-J\left(\mathbf{f}^+\right)
    =&\left\{n^\text{LO}\left[f_1^{\rm HV,LO}-\underset{\mathbf{f}\in \mathcal{S}}{\text{\rm max}}\ f_1^{\rm HV,LO}- \frac{1}{\mu}\left(f_1^{\rm HV,LO}-\underset{\mathbf{f}\in \mathcal{S}}{\text{\rm max}}\ f_1^{\rm HV,LO}+f_1^{\rm HV,HO}-\underset{\mathbf{f}\in \mathcal{S}}{\text{\rm min}}\ f_1^{\rm HV,HO} \right)\right]\right.\nonumber\\
    &+\left.n^\text{HO}\left(f_1^{\rm HV,HO}-\underset{\mathbf{f}\in \mathcal{S}}{\text{\rm min}}\ f_1^{\rm HV,HO}\right)\right\}D_1\left(\phi_1^*\right)\nonumber\\
    &-\left\{n^\text{LO}\left[f_1^{\rm HV,LO}-\underset{\mathbf{f}\in \mathcal{S}}{\text{\rm max}}\ f_1^{\rm HV,LO}-\frac{1}{\mu}\left(f_1^{\rm HV,LO}-\underset{\mathbf{f}\in \mathcal{S}}{\text{\rm max}}\ f_1^{\rm HV,LO}+f_1^{\rm HV,HO}-\underset{\mathbf{f}\in \mathcal{S}}{\text{\rm min}}\ f_1^{\rm HV,HO} \right)\right]\right.\nonumber\\
    &+\left.n^\text{HO}\left(f_1^{\rm HV,HO}-\underset{\mathbf{f}\in \mathcal{S}}{\text{\rm min}}\ f_1^{\rm HV,HO}\right)\right\}D_2\left(\sum_{p\in  \mathcal{P}}\delta^{p}-\phi_1^*\right)\\
    =&\left\{n^\text{LO}\left[f_1^{\rm HV,LO}-\underset{\mathbf{f}\in \mathcal{S}}{\text{\rm max}}\ f_1^{\rm HV,LO}\right.-\frac{1}{\mu}\left(f_1^{\rm HV,LO}-\underset{\mathbf{f}\in \mathcal{S}}{\text{\rm max}}\ f_1^{\rm HV,LO}+f_1^{\rm HV,HO}-\underset{\mathbf{f}\in \mathcal{S}}{\text{\rm min}}\ f_1^{\rm HV,HO} \right)\right]\nonumber\\
    &+\left.n^\text{HO}\left(f_1^{\rm HV,HO}-\underset{\mathbf{f}\in \mathcal{S}}{\text{\rm min}}\ f_1^{\rm HV,HO}\right)\right\} \left(D_1\left(\phi_1^*\right)-D_2\left(\sum_{p\in  \mathcal{P}}\delta^{p}-\phi_1^*\right)\right)\\
    =&\left[n^\text{LO}\left(1-\frac{1}{\mu}\right)\left(f_1^{\rm HV,LO}-\underset{\mathbf{f}\in \mathcal{S}}{\text{\rm max}}\ f_1^{\rm HV,LO}\right)\right.\nonumber\\
    &+\left.\left(n^\text{HO}-\frac{n^\text{LO}}{\mu}\right)\left(f_1^{\rm HV,HO}-\underset{\mathbf{f}\in \mathcal{S}}{\text{\rm min}}\ f_1^{\rm HV,HO} \right)\right] \left(D_1\left(\phi_1^*\right)-D_2\left(\sum_{p\in  \mathcal{P}}\delta^{p}-\phi_1^*\right)\right).
\end{align}
According to Equation~\eqref{eq:f1*}, we have 
\begin{align}
    D_1\left(\phi_1^*\right)+\tau=D_2\left(\sum_{p\in  \mathcal{P}}\delta^{p}-\phi_1^*\right).\nonumber
\end{align}
Since $\tau> 0$, we have 
\begin{align}
    D_1\left(\phi_1^*\right)-D_2\left(\sum_{p\in  \mathcal{P}}\delta^{p}-\phi_1^*\right)< 0.
\end{align}
Also, we have $1-\frac{1}{\mu}< 0$, $n^\text{HO}-\frac{n^\text{LO}}{\mu}>0$, $f_1^{\rm HV,LO}-\underset{\mathbf{f}\in \mathcal{S}}{\text{\rm max}}\ f_1^{\rm HV,LO}\leq 0$ and $f_1^{\rm HV,HO}-\underset{\mathbf{f}\in \mathcal{S}}{\text{\rm min}}\ f_1^{\rm HV,HO}\geq 0$. Therefore, we can conclude that
\begin{align}
    J\left(  \mathbf{f}\right)-J\left(  \mathbf{f}^+\right)\leq 0.\nonumber
\end{align}
To complete the proof, we need to also show that $\mathbf{f}^+\in \mathcal{S}$ exists. Therefore, we aim to prove that there exists some flow $f_1^{+{\rm AV,LO}}$ that satisfies the following conditions:
\begin{align}
    \underset{\mathbf{f}\in \mathcal{S}}{\text{\rm max}}\ f_1^{\rm HV,LO}+\underset{\mathbf{f}\in \mathcal{S}}{\text{\rm min}}\ f_1^{\rm HV,HO}+ \mu f_1^{+{\rm AV,LO}} &=\phi_1^*-\delta^{\rm AV,HO},\\
    0\leq  f_1^{+{\rm AV,LO}} \leq d^{\rm AV,LO}.
\end{align}
Equivalently, we aim to show that 
\begin{align}
\label{eq:f1AVLOex}
    0\leq \phi_1^*-\delta^{\rm AV,HO}-\underset{\mathbf{f}\in \mathcal{S}}{\text{\rm max}}\ f_1^{\rm HV,LO}-\underset{\mathbf{f}\in \mathcal{S}}{\text{\rm min}}\ f_1^{\rm HV,HO} \leq \mu d^{\rm AV,LO}.
\end{align}
We first prove the left-hand-side inequality of Eq. \eqref{eq:f1AVLOex}. Let $\Tilde{f}_1^{\rm HV,HO}$ be any flow of HV,HOs on Lane 1 at any $\mathbf{f}\in\mathcal{S}$ given that the flow of HV,LOs on Lane 1 equals $\underset{\mathbf{f}\in \mathcal{S}}{\text{\rm max}}\ f_1^{\rm HV,LO}$. According to Equation~\eqref{eq:range}, we must have
\begin{align}
    \Tilde{f}_1^{\rm HV,HO}\leq \phi_1^*-\delta^{\rm AV,HO}-\underset{\mathbf{f}\in \mathcal{S}}{\text{\rm max}}\ f_1^{\rm HV,LO}.
\end{align}
Due to $\underset{\mathbf{f}\in \mathcal{S}}{\text{\rm min}}\ f_1^{\rm HV,HO}\leq \Tilde{f}_1^{\rm HV,HO}$, we have
\begin{align}
    \underset{\mathbf{f}\in \mathcal{S}}{\text{\rm min}}\ f_1^{\rm HV,HO}\leq \phi_1^*-\delta^{\rm AV,HO}-\underset{\mathbf{f}\in \mathcal{S}}{\text{\rm max}}\ f_1^{\rm HV,LO},
\end{align}
which proves the left-hand-side inequality of Eq. \eqref{eq:f1AVLOex}. 

We now prove the right-hand-side inequality. Let $\Tilde{f}_1^{\rm HV,LO}$ be any flow of HV,LOs on Lane 1,  at any $\mathbf{f}\in\mathcal{S}$, when the flow of HV,HOs on Lane 1 equals $\underset{\mathbf{f}\in \mathcal{S}}{\text{\rm min}}\ f_1^{\rm HV,HO}$. According to Equation~\eqref{eq:range}, we must have
\begin{align}
    \phi_1^*-\delta^{\rm AV,HO}-\underset{\mathbf{f}\in \mathcal{S}}{\text{\rm min}}\ f_1^{\rm HV,HO}&= \Tilde{f}_1^{\rm HV,LO}+\mu {f}_1^{+\rm AV,LO}.
\end{align}
Since $\Tilde{f}_1^{\rm HV,LO}+\mu {f}_1^{+\rm AV,LO} \leq  \underset{\mathbf{f}\in \mathcal{S}}{\text{\rm max}}\ f_1^{\rm HV,LO}+\mu d^{\rm AV,LO}$, we have
\begin{align}
    \phi_1^*-\delta^{\rm AV,HO}-\underset{\mathbf{f}\in \mathcal{S}}{\text{\rm min}}\ f_1^{\rm HV,HO}\leq  \underset{\mathbf{f}\in \mathcal{S}}{\text{\rm max}}\ f_1^{\rm HV,LO}+\mu d^{\rm AV,LO},
\end{align}
which proves the right-hand-side inequality of Eq. \eqref{eq:f1AVLOex}. Thus the proof is complete.
\end{proof}

\begin{remark} [Implications of Theorem~\ref{thm:characterization}]
Theorem~\ref{thm:characterization} compares the impact on social mobility of vehicles' occupancy and autonomy features, or in other words, the capabilities of AV,LOs and HV,HOs to decrease the total commuter delay. When $\nu^{\rm HV,HO}>\nu^{\rm AV,LO}$, HV,HOs are more capable than AV,LOs in decreasing total commuter delay. Therefore, among the multiple equilibria, the best-case equilibrium that minimizes the total commuter delay happens when we prioritize high-occupancy vehicles instead of autonomous vehicles on toll lane 1. Conversely, when $\nu^{\rm HV,HO} < \nu^{\rm AV,LO}$, HV,HOs are less capable than AV,LOs in decreasing total commuter delay, and therefore, among the multiple equilibria, the best-case equilibrium that minimizes total commuter delay happens when we prioritize autonomous vehicles instead of high-occupancy vehicles on toll lane 1. In all cases, HV,LOs have the smallest mobility degree (recall that $\nu^{\rm HV,LO}=n^\text{LO}$ and~\eqref{eq:def_nu}). Thus the worst-case equilibria that maximize the total commuter delay happen when we prioritize HV,LOs on toll lane 1. 
\end{remark}

In the section, Theorem~\ref{thm:uniqueness} provides the conditions under which there is a unique equilibrium, while Theorem~\ref{thm:characterization} allows us to determine the best and worst case equilibria, in terms of the total commuter delay, when the conditions in Theorem~\ref{thm:uniqueness} do not hold.
To conclude this section, we provide a numerical example to better illustrate these results.

\begin{example}\label{eg:1}
Let $\mathbf{d}=\{d^{\rm AV,HO}=4,\ d^{\rm AV,LO}=3,\ d^{\rm HV,HO}=4,\ d^{\rm HV,LO}=5\}$ (in unit of passengers/minute). Assume the delay functions as BPR functions~\cite{roads1964traffic} in the form:
\begin{align}
\label{eq:BPR}
    D_i(\phi_i)=\theta_i+\gamma_i\left(\frac{\phi_i}{m_i}\right)^{\beta_i},\ \forall i \in \mathcal{I},
\end{align}
with parameters $\mathbf{D}=\{\theta_i=3\ (\text{in minutes}),\ \gamma_i=1,\ \beta_i=1,\ m_i=10\ (\text{in vehicles/minute}):\ i\in \mathcal{I} = \{1,2\}\}$. When $\{n^{\text{LO}}=1,\ n^{\text{HO}}=4,\ \mu =0.5\}$, the lane choice equilibrium always exists and is unique when $\tau\geq 0.7$. Setting $\tau=0.5$, the resulting equilibria form a simplex. The best-case equilibrium in terms of the total commuter delay lies at $\left(f_1^{\rm HV,LO}=0,\ f_1^{\rm HV,HO}=1,\ f_1^{\rm AV,LO}=0\right)$, and the worst equilibrium in terms of the total commuter delay lies at $\left(f_1^{\rm HV,LO}=1,\ f_1^{\rm HV,HO}=0,\ f_1^{\rm AV,LO}=0\right)$. When $\{n^{\text{LO}}=1,\ n^{\text{HO}}=2,\ \mu =0.4\}$, the equilibrium is unique when $\tau\geq 0.74$. Setting $\tau=0.5$,
we have the best equilibrium at 
$\left(f_1^{\rm HV,LO}=0,\ f_1^{\rm HV,HO}=0,\ f_1^{\rm AV,LO}=3\right)$, and the worst equilibrium at $\left(f_1^{\rm HV,LO}=1.2,\ f_1^{\rm HV,HO}=0,\ f_1^{\rm AV,LO}=0\right)$. All the vehicle flows are in units of vehicles/minute. 
It should be noted that the demand and delay parameters are initially specified with specific units to enhance realism. However, the applicability of these examples does not depend on the units chosen. Therefore, in subsequent examples, units will be omitted for simplicity.
\end{example}

\section{Applications in Optimizing Mixed Autonomy Transportation} \label{sec:design}

In this section, we delve into various application domains where the toll lane framework plays a pivotal role in enhancing decision-making and optimization efforts.

\subsection{Designing the uniform toll} \label{sec:toll design}

One important transportation policy problem is to determine an appropriate toll, which ideally induces the resulting equilibrium to approximate a socially optimal one. The toll design problem can be formulated as 
\begin{align}
    \underset{\tau\geq 0}{\text{min}} &\ \ \ J(\mathbf{f})\nonumber\\
    \text{subject to} &\ \ \ \text{Conditions} ~\eqref{eq:2}-~\eqref{eq:eq_def}.\nonumber
\end{align}
Solving optimization problems that incorporate equilibrium constraints is notoriously complex. Nevertheless, leveraging the equilibrium characterization provided in section~\ref{sec:NE}, we introduce a straightforward yet efficient approach below to pinpoint the optimal toll for minimizing the total commuter delay.

Given a specific toll value, as per Theorem~\ref{thm:uniqueness}, the equilibrium, is either singular or resides within the simplex $\mathcal{S}$~\eqref{eq:S}. This simplex can be obtained by solving the equation~\eqref{eq:f1*}. Furthermore, Theorem~\ref{thm:characterization} facilitates the identification of equilibria that represent the best and worst cases in terms of the total commuter delay by examining the contour of $\mathcal{S}$. Consequently, for each candidate toll value, it is feasible to compute the total commuter delay or the extremities of the total commuter delay, transforming the toll design problem into a manageable one-dimensional search task. This search can be efficiently executed using established methodologies like the golden section search~\cite{d6e06843-9f44-3b92-881d-f5d31b55c5e7}, streamlining the process of finding the optimal toll. We further use an example for better illustration.

\begin{figure}[htb]
\centering
 \includegraphics[width=0.5\textwidth]{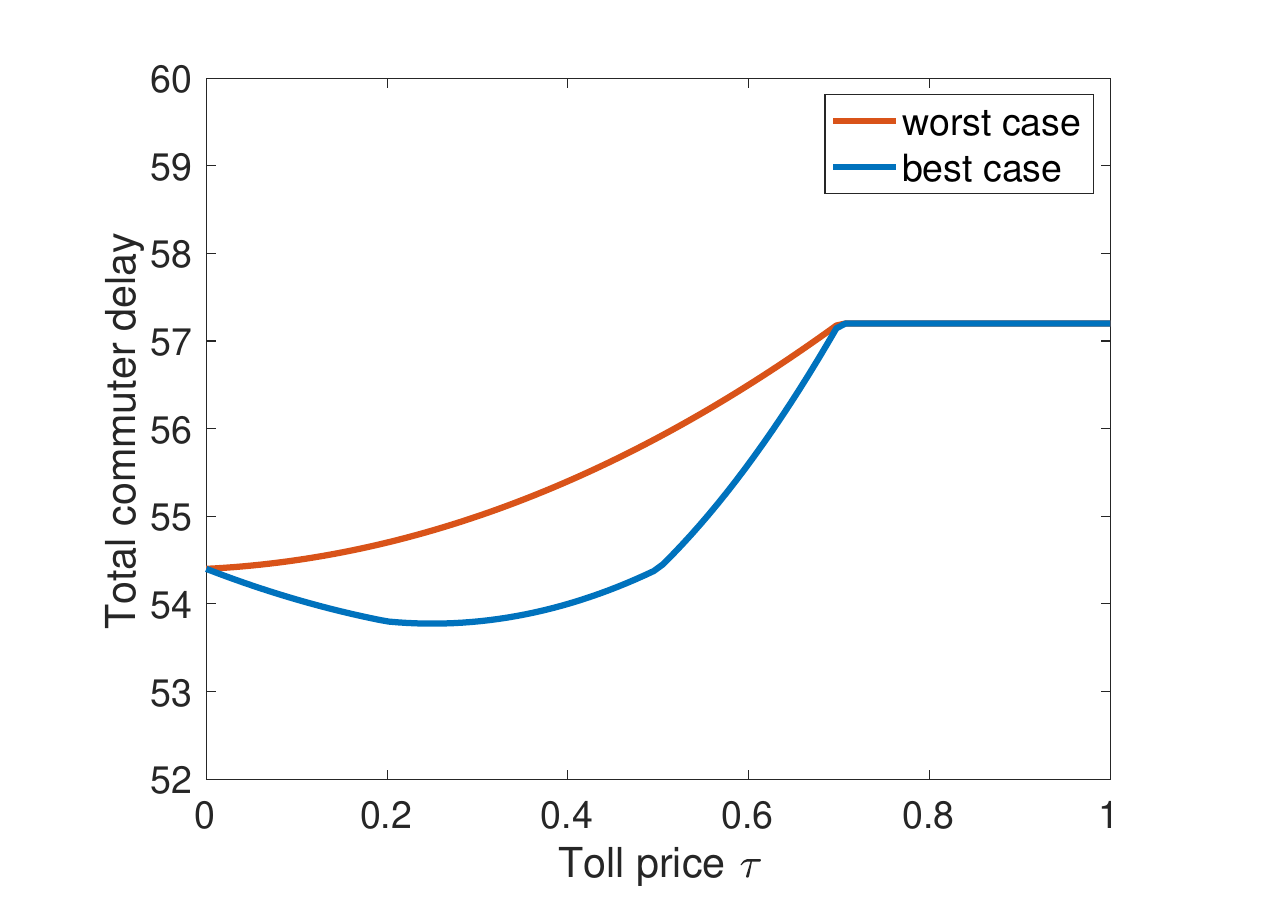}
    \caption{The best/worst-case total commuter delay versus different toll values,  when a uniform toll is imposed on all vehicles traveling on lane 1, except for autonomous vehicles with high occupancy (AV,HOs). We can set $\tau = 0$ to minimize the worst-case total commuter delay or set the toll rate at approximately $\tau = 0.25$ to minimize the best-case total commuter delay.}
    \label{fig:toll}
\end{figure}

\begin{example}\label{eg:toll}
Consider again Example \ref{eg:1} where \{$n^{\text{LO}}=1,\ n^{\text{HO}}=4,\ \mu = 0.5$\}.
The best/worst case total commuter delay at different toll values are characterized and shown in Figure~\ref{fig:toll}. Per the figure, the worst-case total commuter delay increases as the toll increases, until the equilibrium becomes unique when $\tau \ge 0.7$. In contrast, the best-case total commuter delay first decreases with the increasing toll, until it reaches a global minimum at $\tau = 0.25$, and then increases. 
Based on the desired outcome, the selection of the toll rate can vary: setting $\tau = 0$ is optimal for minimizing the worst-case total commuter delay, whereas a toll rate of approximately $\tau = 0.25$ is preferable for minimizing the best-case total commuter delay.
\end{example}

\subsection{Designing the occupancy threshold} \label{sec:n design}


The occupancy threshold, $n$, is designated as the minimum number of commuters per vehicle required for it to be classified as a high-occupancy vehicle, thereby qualifying for associated benefits. Determining an appropriate value for $n$ is a common and non-trivial task.
While policymakers may be inclined to raise $n$ to foster higher occupancy rates within vehicles, excessively increasing this threshold beyond $n = 2$ complicates and elevates the costs of engaging in carpooling activities, therefore, an increased $n$ could paradoxically decrease the overall demand for carpooling.

Our methodology extends to determining the optimal passenger occupancy threshold, $n$. This approach parallels the process used for finding the optimal uniform toll, featuring a straightforward and efficient algorithm. For any candidate $n$ value, we can estimate the corresponding high-occupancy and low-occupancy vehicle demands, using any existing demand models, along with the average high occupancy ($n^{\text{HO}}$) and the average low occupancy ($n^{\text{LO}}$). Thus we can identify either a unique total commuter delay for a resulting singular equilibrium (Theorem~\ref{thm:uniqueness}), or to pinpoint the best and worst case total commuter delays within the equilibrium simplex $\mathcal{S}$~\eqref{eq:S} (using Theorem~\ref{thm:characterization}).
Consequently, optimizing the passenger occupancy threshold transforms into a manageable one-dimensional search. To provide a clearer understanding, we will illustrate this method with a specific example.

\begin{figure}[hbt]
\centering
 \includegraphics[width=0.5\textwidth]{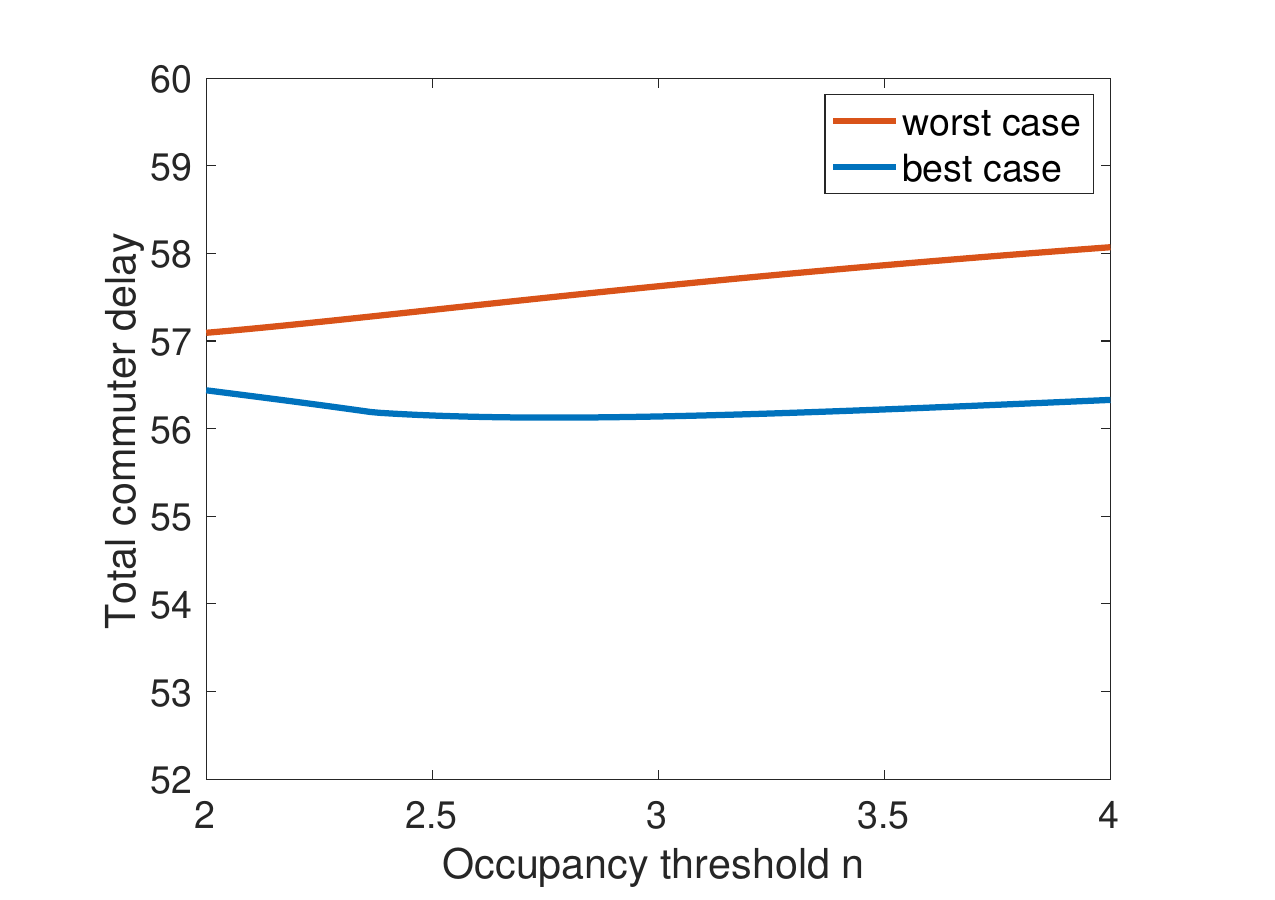}
    \caption{The best/worst case total commuter delays versus different values of the occupancy threshold $n$ in Example~\ref{eg:n}. In this example, increasing the occupancy threshold $n$ does not significantly decrease the best-case total commuter delay, whereas the worst-case total commuter delay increases evidently, consequently, enhancing the occupancy threshold may not be an effective strategy.}\label{fig:thresh_p}
\end{figure}

\begin{example}\label{eg:n} 

We employ a basic, or "vanilla" demand model as an illustrative example, however, it is important to note that our methodology is adaptable to other demand models.

We start by assuming that commuter demands for human-driven vehicles (denoted by $d^{\rm HV}$) and autonomous vehicles (denoted by $d^{\rm AV}$) remain inelastic. For an occupancy threshold of $n\geq 2$, we consider the probability of carpooling (to achieve or surpass this threshold) as $u(n)\in [0,1]$. This probability is assumed to be the same for both human-driven and autonomous vehicles. The function $u(\cdot)$, moreover, is designed to be non-increasing, which mirrors the observed decline in carpooling willingness as the occupancy requirement increases. We assume vehicles either commit to carpooling, achieving the high-occupancy vehicle status with occupancy $n$, or opt to remain as single-occupant vehicles. Thus, we have 
\begin{align}
\label{eq:uprobn}
    d^{\rm HV,HO} =d^{\rm HV}u(n),& \ \ \ \  d^{\rm HV,LO} =d^{\rm HV}(1-u(n)),\nonumber\\
    d^{\rm AV,HO} =d^{\rm AV}u(n),& \ \ \ \  d^{\rm AV,LO} =d^{\rm AV}(1-u(n)),\\
    n^{\text{HO}}=n,& \ \ \ \ n^{\text{LO}}=1. \nonumber
\end{align}
To find the optimal passenger occupancy threshold $n$, we need to solve the optimization problem:
\begin{align}
    \underset{n\geq 2}{\text{min}} &\ \ \ J(\mathbf{f})\nonumber\\
    \text{subject to} &\ \ \ \text{Conditions} ~\eqref{eq:2}-~\eqref{eq:eq_def}\:\: \text{and} \:\: \eqref{eq:uprobn}.\nonumber
\end{align}

Let $\{d^{\rm AV}=7,\ d^{\rm HV}=9,\ \mu =0.5,\ \tau =0.5\}$ and the delay functions are given by the BPR function in Eq.~\eqref{eq:BPR}, with parameters $\mathbf{D}=\{\theta_i=3,\ \gamma_i=1,\ \beta_i=1,\ m_i=10:\ i\in \mathcal{I}\}$. The probability of a commuter to carpool is given by $u(n)=\frac{1}{n}$ for $n\in[2,4]$. The corresponding best/worst case total commuter delays under each value of $n$ are shown in Figure~\ref{fig:thresh_p}. For this example, increasing the occupancy threshold $n$ does not significantly decrease the best-case total commuter delay, whereas the worst-case total commuter delay increases evidently. Consequently, enhancing the occupancy threshold may not be an effective strategy in this example.

\end{example}

\subsection{Designing the lane policy} \label{sec:policy design}


This section delves into the comparison between lanes designated for high-occupancy vehicles and those reserved for autonomous vehicles. Traditionally, transit policies have allowed high-occupancy vehicles to use a dedicated HOV lane without any toll, while permitting other vehicles access to this lane for a toll, a practice known as the dedicated High-Occupancy Vehicles Lanes (HOVL).
With the imminent advancement of autonomous and connected driving technologies, it becomes pertinent to assess the impact of maintaining HOVL versus adopting lanes that allow autonomous vehicles unfettered access (while charging non-autonomous vehicles a toll for lane usage), hereafter referred to as the Dedicated Lanes for Autonomy (DLA).

Leveraging the modeling framework outlined in the preceding sections, we can analyze and compare the two distinct lane policies: the High-Occupancy Vehicle Lane (HOVL) policy and the Dedicated Lane for Autonomy (DLA) policy. To model the HOVL policy within our framework, we assign all high-occupancy vehicle traffic to lane 1, i.e., we let $f_1^{\rm HV,HO}=d_\text{v}^{\rm HV,HO}$.
In contrast, the DLA policy is modeled by letting $f_1^{\rm AV,LO}=d_\text{v}^{\rm AV,LO}$, thereby ensuring all autonomous vehicles have dedicated access to Lane 1. 
Then we use theorems~\ref{thm:uniqueness} and~\ref{thm:characterization} to characterize resulting singular or multiple equilibria for comparison. An example is shown below for better illustration.

\begin{figure}[hbt]
\centering
 \includegraphics[width=0.5\textwidth]{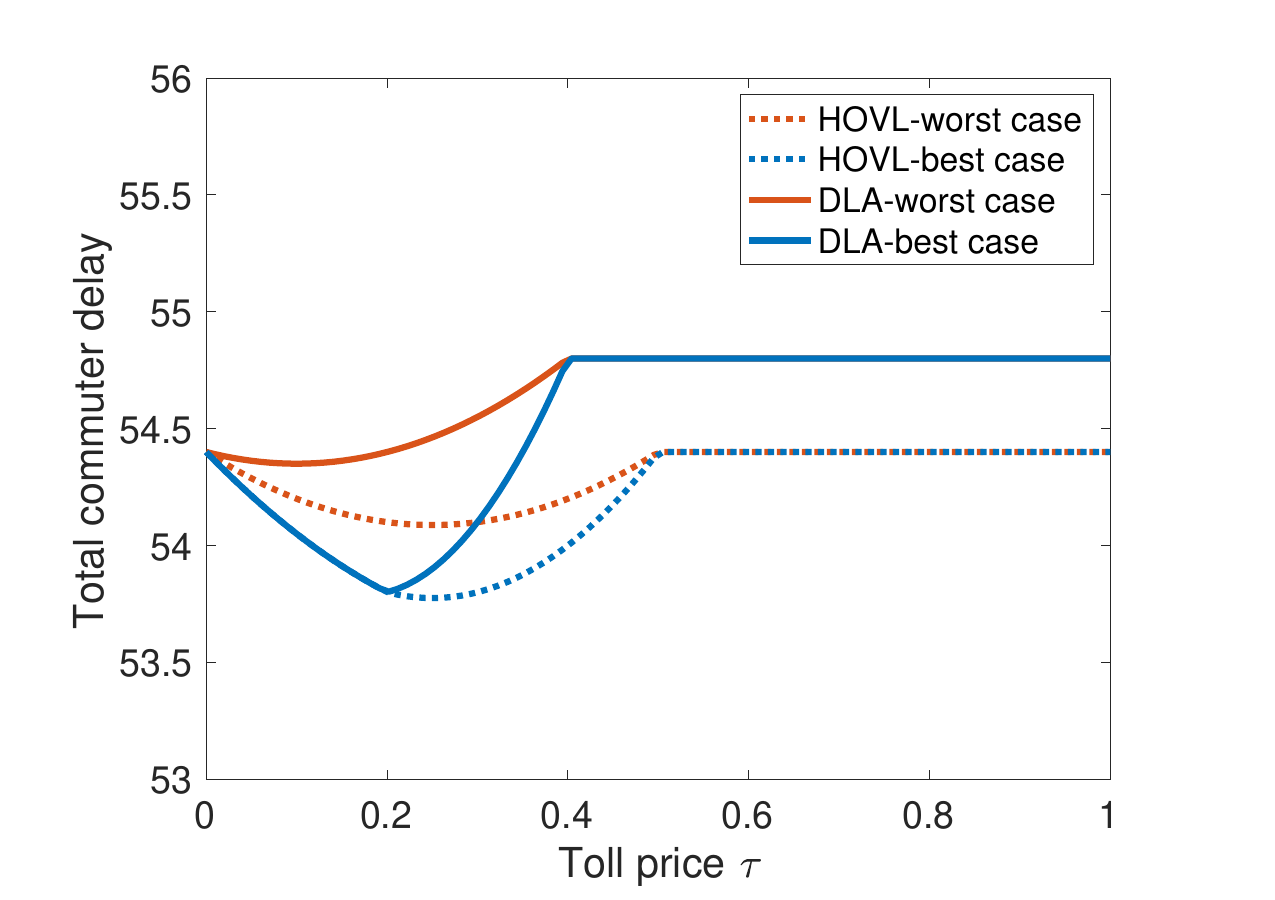}
    \caption{The best/worst case commuter total commuter delay versus toll rates under different lane policies in Example~\ref{eg:policy}. For the specific highway configuration and any toll value listed, the HOV Lane (HOVL)  policy outperforms the Dedicated Lane for Autonomy (DLA) policy and indicates a better strategy. }\label{fig:policy}
\end{figure}

\begin{example}\label{eg:policy}
Let $\{d^{\rm AV,HO}=4,\ d^{\rm AV,LO}=3,\ d^{\rm HV,HO}=4,\ d^{\rm HV,LO}=5,\ n^\text{HO}=4,\ n^\text{LO}=1,\ \mu =0.5\}$. Assume the delay functions as BPR functions with parameters $\mathbf{D}=\{\theta_i=3,\ \gamma_i=1,\ \beta_i=1,\ m_i=10:\ i\in \mathcal{I}\}$. The best/worst case total commuter delay for each of the two policies is shown in Figure~\ref{fig:policy}. For the specific highway configuration and any toll value listed, the HOV Lane (HOVL)  policy outperforms the Dedicated Lane for Autonomy (DLA) policy, and indicates a better strategy.
\end{example}

\section{Strategic Toll Differentiation for Equilibrium Optimization} \label{sec:diff_toll}

Acknowledging the frequent occurrence of non-unique lane choice equilibria, this section proposes a targeted strategy. We assign differentiated toll rates to each of the three decision-making vehicle classes, ensuring that each class faces a unique toll rate. This approach is designed to strategically influence the traffic equilibria towards a more favorable outcome, thereby improving the overall mobility.

Before diving into the specifics of the differentiated toll strategy, we take a step back to reinforce our model within the context of heterogeneous tolls. We let $\bm{\tau} \in \mathbb{R}_+^{3}$ be the heterogeneous toll vector defined as
\begin{align}
\label{eq:tauvec}
    \bm{\tau} :=& \left(\tau^{\rm HV,LO},\ \tau^{\rm HV,HO},\ \tau^{\rm AV,LO}\right),\\
    &\ \ \tau^p > 0, \ \ \forall p \in \bar{\mathcal{P}}.
\end{align}
Most importantly, every decision-making vehicle class has a unique toll rate, i.e., 
\begin{align}
   \tau^p \ne \tau^q,\ \ \forall p \ne q\in \bar{\mathcal{P}}.
\end{align}
Correspondingly, we let $J(\bm{\tau})$ denote the total commuter delay of the equilibrium induced by the heterogeneous toll vector  $\bm{\tau}$. 

Let $C_1^{p}$ be the travel cost for vehicles of class $ p\in \bar{\mathcal{P}}$ on lane 1, and we have

\begin{align}
    C_1^p(\mathbf{f}) &=D_1(\phi_1)+\tau^p,\:\:\: \forall p\in \bar{\mathcal{P}}. 
    \label{eq:tL1}
\end{align}
In contrast, the travel cost for all vehicles on lane 2 remains consistent and is equivalent to the delay experienced:
\begin{align}
    C_2(\mathbf{f}) &=D_2(\phi_2).\label{eq:tl4}
\end{align}
Therefore, the lane choice equilibrium of the decision-making vehicles under differentiated tolls can then be modeled as a Wardrop equilibrium as follows.

\begin{definition}[Lane choice equilibrium under heterogeneous tolls]\label{def:wdp_diff}
For a tolled segment of highway $G=\left(\mathbf{D},\mathbf{d},n^{\text{LO}}, n^{\text{HO}},\mu,\bm \tau\right)$ under differentiated tolls, a feasible flow distribution vector $\mathbf{f}$ is a lane choice equilibrium if and only if, for all  vehicles of class $ p\in \bar{\mathcal{P}}$,
    \begin{align}
    f_1^{p} (C_1^{p}(\mathbf{f}) - C_2(\mathbf{f}) &\leq 0, \nonumber \label{eq:eq_def_more} \\
    f_2^{p} (C_2(\mathbf{f}) - C_1^{p}(\mathbf{f})) &\leq 0.
    \end{align}
\end{definition}

When tolls vary across different vehicle classes, the optimization problem described in Section~\ref{sec:toll design} evolves into a nontrivial bi-level multi-dimensional optimization problem with equilibrium constraints. Such problems could be addressed through iterative optimization techniques. However, these methods may suffer from prolonged convergence times, and for delay configurations that are non-convex, convergence cannot always be assured. To circumvent these issues, we introduce a novel approach that reduces total commuter delay through toll differentiation.

Our method unfolds in two sequential steps. It commences by identifying an optimal uniform toll that minimizes the best-case total commuter delay, as outlined in Section~\ref{sec:toll design} (\textbf{Step 1}). If this optimal uniform toll is strictly positive and the system presents multiple equilibria, we then proceed to establish a heterogeneous toll vector that steers the system towards the best-case equilibrium found at the optimal uniform toll. The specific heterogeneous toll vector, $\bm{\tau}$, is identified based on the criteria set forth in the ensuing proposition (\textbf{Step 2}). This approach allows for a significant reduction in total commuter delay without relying on the tentative convergence of an iterative algorithm.

\begin{proposition}[\bf{Toll Differentiation}] \label{prop:diff_tolls}
Consider a tolled segment of highway $G=\left(\mathbf{D},\mathbf{d},n^{\text{LO}}, n^{\text{HO}},\mu,\tau^*\right)$, where $\tau^*>0$ is a predetermined optimal uniform toll that induces non-unique equilibria. Let $\phi_1^*$ be the effective vehicle flow in lane 1 at the equilibria and let $J^*(\tau^*)$ be the best-case total commuter delay under the uniform toll $\tau^*$. A heterogeneous toll vector $\bm{\tau}$~\eqref{eq:tauvec} can be determined as follows to induce a unique lane choice equilibrium that results in $J\left(\bm{\tau}\right) = J^*\left(\tau^*\right)$.

\begin{description}
    \item[1)] If $\frac{n^\text{HO}}{n^\text{LO}}\leq \frac{1}{\mu}$,    or equivalently, $\nu^{\rm HV,HO} \leq \nu ^{\rm AV,LO}$,\\
    
    \begin{description}
        \item[a)] if $\phi_1^*\leq \delta^{\rm AV,LO}+\delta^{\rm AV,HO}$, then set $\bm{\tau}=\left(\tau_+,\ \tau_+,\ \tau^*\right)$;\\[0.25em]
        \item[b)] if $\delta^{\rm AV,LO}+\delta^{\rm AV,HO}< \phi_1^*\leq \delta^{\rm HV,HO}+\delta^{\rm AV,LO}+\delta^{\rm AV,HO} $, then set                     $\bm{\tau}=\left(\tau_+,\tau^*,\tau_-\right)$;\\[0.25em]
        \item[c)] if $\phi_1^*>\delta^{\rm HV,HO}+\delta^{\rm AV,LO}+\delta^{\rm AV,HO} $, then set $\bm{\tau}=\left(\tau^*,\tau_-,\tau_-\right)$.\\

    \end{description}
    \item[2)] Conversely, if $\frac{n^\text{HO}}{n^\text{LO}}> \frac{1}{\mu}$, and equivalently, $\nu^{\rm HV,HO} > \nu ^{\rm AV,LO}$,\\
    
    \begin{description}
        \item[a)] if $\phi_1^*\leq \delta^{\rm HV,HO}+\delta^{\rm AV,HO} $, then set $\bm{\tau}=\left(\tau_+,\tau^*,\tau_+\right)$;\\[0.25em]
        \item[b)] if $\delta^{\rm HV,HO}+\delta^{\rm AV,HO}< \phi_1^*\leq \delta^{\rm AV,LO}+\delta^{\rm HV,HO}+\delta^{\rm AV,HO} $, then set $\bm{\tau}=\left(\tau_+,\tau_-,\tau^*\right)$;\\[0.25em]
        \item[c)] if $\phi_1^*>\delta^{\rm AV,LO}+\delta^{\rm HV,HO}+\delta^{\rm AV,HO} $, then set $\bm{\tau}=\left(\tau^*,\tau_-,\tau_-\right)$.
    \end{description}
\end{description}
Notice that $\tau_-$ and $\tau_+$ can be any toll values that satisfy 
\begin{align}
    0<\tau_-&< \tau^*,\\
    \tau_+&> \tau^*.
\end{align}
\end{proposition}

\begin{proof}
We sketch the proof for Proposition~\ref{prop:diff_tolls}, which is linked to the underpinnings of Theorem~\ref{thm:uniqueness} and~\ref{thm:characterization}. We give the detailed explanation for the first sub-case \textbf{a)} when $\frac{n^\text{HO}}{n^\text{LO}}\leq \frac{1}{\mu}$. According to Theorem~\ref{thm:characterization}, when $\frac{n^\text{HO}}{n^\text{LO}}\leq \frac{1}{\mu}$, i.e. $\nu ^{\rm AV,LO} \geq \nu^{\rm HV,HO}$, we should prioritize AV,LOs on lane 1 so the total commuter delay can be minimized among the multiple lane choice equilibria. When $\phi_1^*\leq \delta^{\rm AV,LO}+\delta^{\rm AV,HO}$, the best-case equilibrium under the optimal uniform toll would be $\mathbf{f}_1 = (0,0,\frac{\phi_1^*-\delta^{\rm AV,HO}}{\mu})$. Further, one can check by Definition~\ref{def:wdp_diff} that $\mathbf{f}_1 =(0,0,\frac{\phi_1^*-\delta^{\rm AV,HO}}{\mu})$ is an equilibrium and is the unique equilibrium when tolls are selected as $\bm{\tau}=\left(\tau_+,\ \tau_+,\ \tau^*\right)$. The other cases can be proved following a similar logic, and thus are omitted.
\end{proof}

\begin{remark} [Interpretation of Proposition~\ref{prop:diff_tolls}]
    The essence of this toll differentiation proposition is to first identify the best-case equilibrium among all equilibria under the optimal homogeneous toll $\tau^*$ , and then assign the optimal homogeneous toll  $\tau^*$ to the vehicle class that uses both lanes under the best-case equilibrium. A toll higher than $\tau^*$ is charged to the vehicle classes with a smaller mobility degree (i.e., they are less capable in increasing social mobility) so that they are incentivized to stay out of the toll lane, and a toll lower than $\tau^*$ to vehicle classes with a larger mobility degree (i.e., they are more capable in increasing social mobility) so that they are incentivized to use the toll lane.
\end{remark}

\section{Evaluating Framework Resilience to Toll Non-Compliance} \label{sec:misbehavior}

The preceding section illustrates how setting differentiated tolls for various vehicle classes accessing the toll lane can substantially reduce the total commuter delay on a freeway segment. Nonetheless, the practical application of differentiated tolls could be compromised by instances of vehicular non-compliance, such as toll evasion. A common scenario arises when toll collection systems depend on signals from vehicle transponders, which commuters themselves often set, to indicate the total number of commuters per vehicle and therefore distinguish between high-occupancy and low-occupancy vehicles. Due to sporadic verification of these signals by toll agencies and inconsistent enforcement of penalties, there is a possibility that some low-occupancy vehicles might dishonestly emit high-occupancy transponder signals, thus wrongfully benefiting from the reduced toll rates designated for high-occupancy vehicles.

In this section, we aim to examine the repercussions of such deceitful practices on the efficacy of the transportation network. As in the previous section, we allow autonomous vehicles with high occupancy (AV,HOs) to travel on the toll lane freely whereas differentiated nonzero tolls $\boldsymbol \tau$ as defined in~\eqref{eq:tauvec} are charged to the other three decision-making vehicle classes. We assume that all decision-making vehicle classes could potentially engage in this misbehavior. We denote the decision-making vehicles that decide to travel on lane 1 without paying their prescribed toll as {\bf misbehaving vehicles}, whereas the other decision-making vehicles as {\bf honest vehicles}. Misbehaving vehicles choose to travel on lane 1, without paying their assigned toll,  at the risk of being caught by law enforcement and paying a fine.

\begin{remark}[Comparison of misbehaving vehicles and honest vehicles]
   We aim to highlight a key distinction for our readers: while honest vehicles meticulously assess the costs associated with traveling on both lanes, selecting the one that minimizes their cost, misbehaving vehicles deliberately choose to travel on lane 1, either without paying a toll or by paying a reduced toll. Whether misbehaving vehicles have conducted a cost evaluation before the misbehavior is beyond the scope of this analysis. Instead, we consider the observed proportions of such vehicles as fixed and given for the purposes of our study.
\end{remark}

Before delving into the analysis, we first outline key notations updated to account for misbehaving vehicles. For decision-making vehicles of class $p\in \Bar{\mathcal{P}}$, we let $\alpha_{\rm m}^{p} \in [0,1]$ indicate the proportion of misbehaving vehicles among the vehicles of class $p$, and the misbehaving proportions are collected in the vector $\boldsymbol{\alpha}_{\rm m} \in \mathbb{R}_+^3$ as
\begin{eqnarray}
\label{eq:alphavec}
\boldsymbol{\alpha}_{\rm m} := \left (\alpha_{\rm m}^{\rm HV,LO}, \alpha_{\rm m}^{\rm HV,HO}, \, \alpha_{\rm m}^{\rm AV,LO} \right ).
\end{eqnarray}
Unlike misbehaving vehicles that exploit access to toll lane 1 without proper qualification, honest vehicles meticulously assess their costs for both lanes before deciding whether to pay the toll. We denote the flow distribution vector of {\bf honest} vehicles as $\hat{\mathbf{f}}:= \left(\hat{\mathbf{f}}_1, \hat{\mathbf{f}}_2\right)$, 
where for lane $i\in \mathcal{I}$,
\begin{align}
    \hat{\mathbf{f}}_i:= \left(\hat{f}_i^{\rm HV,LO},\ \hat{f}_i^{\rm HV,HO},\ \hat{f}_i^{\rm AV,LO}\right).
\end{align}
Therefore, with consideration of misbehaving vehicles, a feasible and meaningful flow distribution vector of {\bf honest} vehicles must satisfy
\begin{align}
\sum_{i\in \mathcal{I}}\hat{f}_i^{\rm HV,LO} &=d_{\text{v}}^{\rm HV,LO}(1-\alpha_{\rm m}^{\rm HV,LO}),\nonumber\\
 \sum_{i\in \mathcal{I}}\hat{f}_i^{\rm HV,HO} &=d_{\text{v}}^{\rm HV,HO}(1-\alpha_{\rm m}^{\rm HV,HO}),\label{eq:m2}\\
 \sum_{i\in \mathcal{I}}\hat{f}_i^{\rm AV,LO} &=d_{\text{v}}^{\rm AV,LO}(1-\alpha_{\rm m}^{\rm AV,LO}),\nonumber\\
    \hat{f}_i^{p}&\geq 0,\ \ \forall p\in \Bar{\mathcal{P}}.\nonumber
\end{align}

We also define $\hat{\phi}_i$ as the effective vehicle flow on lane $i\in \mathcal{I}$, accounting for  misbehaving vehicles:
\begin{align}
    \hat{\phi}_1 &:=\hat{f}_1^{\rm HV,LO}+\hat{f}_1^{\rm HV,HO}+\mu  \hat{f}_1^{\rm AV,LO}+\delta^{\rm AV,HO}+\sum_{p\in \Bar{\mathcal{P}}}\delta^{p}\alpha_{\rm m}^{p},\label{eq:m_f1}\\
\hat{\phi}_2 &:=\hat{f}_2^{\rm HV,LO}+\hat{f}_2^{\rm HV,HO}+\mu \hat{f}_2^{\rm AV,LO}, \nonumber
\end{align}
where we remind readers that $\delta^p$ are the inelastic effective vehicle demands as defined in Eq.~\eqref{eq:effdem}. We have
\begin{align}
    \hat{\phi}_1&\in \left[\delta^{\rm AV,HO}+\sum_{p\in \Bar{\mathcal{P}}}\delta^{p}\alpha_{\rm m}^{p}, \sum_{p\in \mathcal{P}}\delta^{p}\right],\\
    \hat{\phi}_2&\in \left[0, \sum_{p\in \Bar{\mathcal{P}}}\delta^{p}(1-\alpha_{\rm m}^{p})\right],
\end{align}
where $\sum_{p\in \Bar{\mathcal{P}}}\delta^{p}\alpha_{\rm m}^{p}$ represents the effective total demand of misbehaving vehicles. Thus, the travel delay $D_i$ for lane $i\in \mathcal{I}$, is then a continuous and increasing function of $\hat{\phi}_i$.
As in the previous section, let $\boldsymbol{\tau}$ be the vector of differentiated tolls, as defined in Eq.~\eqref{eq:tauvec}. Therefore, the travel cost experienced by {\bf honest} vehicles of class $p\in \bar{\mathcal{P}}$, respectively traveling on lane 1 and 2 is  given by 
\begin{align}
    C_1^{p}(\hat{\mathbf{f}}) &=D_1(\hat{\phi}_1)+\tau^{p},\label{eq:tL1}\\
    C_2(\hat{\mathbf{f}}) &=D_2(\hat{\phi}_2).\label{eq:tl4}
\end{align}

Now we are ready to update the definition for the lane choice equilibrium. Consider a tolled segment of highway $G=\left(\mathbf{D},\mathbf{d},n^{\text{LO}}, n^{\text{HO}},\mu,\bm \tau,\boldsymbol{\alpha}_{\rm m}\right)$ with the delay models $\mathbf{D}$, commuter demands $\mathbf{d}$, toll prices $\bm \tau$, and misbehavior proportions $\boldsymbol{\alpha}_{\rm m}$. The lane choice equilibrium for honest vehicles can be modeled as follows.

\begin{definition}[Lane choice equilibria considering vehicle misbehavior]\label{def:wdp_more}
For a tolled segment of highway $G=\left(\mathbf{D},\mathbf{d},n^{\text{LO}}, n^{\text{HO}},\mu,\bm \tau,\boldsymbol{\alpha}_{\rm m}\right)$ with differentiated tolls and misbehaving vehicles, a feasible honest vehicle flow distribution vector $\hat{\mathbf{f}}$ is a lane choice equilibrium if and only if, for any vehicle class  $p \in  \Bar{\mathcal{P}}$, we have
    \begin{align}
    \label{eq:eq_def_more}
    \hat{f}_1^{p} (C_1^{p}(\hat{\mathbf{f}}) - C_2(\hat{\mathbf{f}})) &\leq 0 ,\\
    \hat{f}_2^{p} (C_2(\hat{\mathbf{f}}) - C_1^{p}(\hat{\mathbf{f}}))&\leq 0. \nonumber
    \end{align}
\end{definition}

Notice that, in the corner case where there exists no vehicle misbehavior, i.e., $\alpha_{\rm m}^{p}=0$ for all $p\in \bar{\mathcal{P}}$, the lane choice equilibrium in Definition~\ref{def:wdp_more} will be the same as the lane choice equilibrium in Definition~\ref{def:wdp_diff}. To characterize the efficiency of the equilibria, we also update the total commuter delay as
\begin{align}\label{eq:social_cost_more}
   J(\hat{\mathbf{f}}) =&
   \left[ n^\text{LO}\left( \hat{f}_1^{\rm HV,LO}+d_\text{v}^{\rm HV,LO}\alpha_{\rm m}^{\rm HV,LO}+ \hat{f}_1^{\rm AV,LO}+d_\text{v}^{\rm AV,LO}\alpha_{\rm m}^{\rm AV,LO} \right)\right.\nonumber\\
   & + \left. n^\text{HO}\left( \hat{f}_1^{\rm HV,HO}+d_\text{v}^{\rm HV,HO}\alpha_{\rm m}^{\rm HV,HO}+d_\text{v}^{\rm AV,HO}\right)\right] D_1( \hat{\phi}_1)\\
& + \left[ n^\text{LO}\left( \hat{f}_2^{\rm HV,LO}+ \hat{f}_2^{\rm AV,LO}\right)+n^\text{HO}\hat{f}_2^{\rm HV,HO} \right] D_2( \hat{\phi}_2) \nonumber
\end{align}

Investigation into vehicle misbehavior involves assessing its potential to degrade traffic conditions, particularly on toll lane 1, where unauthorized vehicles might contribute to excessive demand. This misconduct not only threatens the efficiency of the designated toll lane but could also ripple through the entire transportation system, undermining its overall functionality. To address these concerns, our analysis commences with an exploration of how such misbehavior impacts lane delays, articulated through the forthcoming theorem:

\begin{theorem} [Lane delays can be resilient to vehicle misbehavior]\label{thm:lane_delay_robust}
Consider a tolled segment of highway $G=\left(\mathbf{D},\mathbf{d},n^{\text{LO}}, n^{\text{HO}},\mu,\bm \tau,\boldsymbol{\alpha}_{\rm m}\right)$ with differentiated tolls and potential vehicle misbehavior. The lane delays at the lane choice equilibrium $\hat{\mathbf{f}}$ (by Definition~\ref{def:wdp_more}), will remain the same as the lane delays at the lane choice equilibrium $\mathbf{f}$ without any vehicle misbehavior (by Definition \ref{def:wdp_diff}), if both of the following two conditions hold:
\begin{description}
    \item[1)] At the lane choice equilibrium $\mathbf{f}$ without any vehicle misbehavior, there exists a class of vehicles  $g\in \Bar{\mathcal{P}}$ that travel on both lanes. Or equivalently, assuming no vehicle misbehavior, there exists a class of vehicles  $g\in \Bar{\mathcal{P}}$ and an effective vehicle flow $\phi_1^*$ that satisfy
    \begin{align}
         D_1(\phi_1^*)+\tau^{g}&=  D_2\left(\sum_{p\in \mathcal{P}}\delta^{p}-\phi_1^*\right),\label{eq:mis1}\\
         \phi_1 &\in \left[\delta^{\rm AV,HO},\sum_{p\in \mathcal{P}}\delta^{p}\right],
    \end{align}
    where $\tau^g$ is the assigned toll to vehicle class $g$.
    \item[2)] Misbehaving vehicle proportions $\boldsymbol{\alpha}_{\rm m}$ satisfy
    \begin{align}
    \label{eq:th3-alpha_bound}
        \sum_{p\in \Bar{\mathcal{P}}\setminus \mathcal{Q}^{g}_{-}}\delta^{p}\alpha_{\rm m}^{p}\leq \phi_1^{*} - \delta^{\rm AV,HO}- \sum_{q\in \mathcal{Q}^{g}_{-}}\delta^{q},
    \end{align}
   where the set  $\mathcal{Q}^{g}_{-}\subset \Bar{\mathcal{P}}:=\{p\in \Bar{\mathcal{P}}:0<\tau^p<\tau^g \} $ contains all vehicle classes that are assigned a toll higher than zero but lower than 
   $\tau^g$.
\end{description}

\end{theorem}

\begin{proof}
We prove by contradiction that, under conditions~\eqref{eq:mis1}-\eqref{eq:th3-alpha_bound}, the effective total flow on lane 1 under two equilibria remain the same, i.e., $\hat{\phi}_1=\phi_1^*$ and as a consequence, the lane delays remain the same. 

Assume first that $\hat{\phi}_1>\phi_1^*$. Since $C_1^g(\phi_1)$ are increasing functions of $\phi_1$ and $C_2^{g}(\phi_1)$ are decreasing functions of $\phi_1$, we conclude that it must be the case that $C_1^{g}(\hat{\phi}_1)>C_1^{g}(\phi_1^*)$ and $C_2^{g}(\hat{\phi}_1)<C_2^{g}(\phi_1^*)$. According to condition~\eqref{eq:mis1}, since  $C_2^{g}(\hat{\phi}_1)<C_1^{g}(\hat{\phi}_1)$, all class $g$ vehicles must be traveling  on lane 2. Correspondingly, all the other classes of vehicles, which are not misbehaving and whose assigned toll on lane 1 is higher than or equal to the toll assigned to vehicles of class $g$, must also be traveling on lane 2. Further, considering some vehicles belonging to $\mathcal{Q}_-^g$ could also be traveling on lane 2, at the equilibrium $\hat{\mathbf{f}}$ described by Eq.~\eqref{eq:eq_def_more}, we have 
\begin{eqnarray}\label{eq:phi1lb}
\hat{\phi_1}\leq  \sum_{p\in \Bar{\mathcal{P}}\setminus \mathcal{Q}^{g}_{-}}\delta^{p}\alpha_{\rm m}^{p}+\delta^{\rm AV,HO}+\sum_{q\in \mathcal{Q}_-^g}\delta^{q}.
\end{eqnarray}
However, with condition~\eqref{eq:th3-alpha_bound}, our conclusion \eqref{eq:phi1lb} implies that $\hat{\phi_1}\leq \phi_1^*$, which contradicts our assumption. 

Next, assume that $\hat{\phi}_1<\phi_1^*$ and, as a consequence,  $C_1^{g}(\hat{\phi}_1)<C_1^{g}(\phi_1^*)$ and $C_2^{g}(\hat{\phi}_1)>C_2^{g}(\phi_1^*)$. According to condition~\eqref{eq:mis1}, we have $C_2^{g}(\hat{\phi}_1)>C_1^{g}(\hat{\phi}_1)$. Thus, all honest vehicles of class $g$ must be traveling on lane 1. Correspondingly, all the other honest vehicles whose toll is lower than $\tau^g$ must also be traveling on lane 1. Further considering the misbehaving vehicles traveling on lane 1, at the equilibrium $\hat{\mathbf{f}}$, we have 
\begin{eqnarray}\label{eq:404}
\hat{\phi_1}\geq \delta^{\rm AV,HO}+\sum_{q\in \mathcal{Q}_-^g}\delta^{q}(1-\alpha_{\rm m}^q)+\sum_{q\in \mathcal{Q}_-^g}\delta^{q}\alpha_{\rm m}^q+\delta^g(1-\alpha_{\rm m}^g)+ \delta^g\alpha_{\rm m}^g.
\end{eqnarray}
However, at equilibrium $\mathbf{f}$ described in \eqref{eq:mis1} when there are no misbehaving vehicles, since vehicles of class $g$ are traveling on both lanes, we have
\begin{eqnarray}\label{eq:040}
\phi_1^*\leq \delta^{\rm AV,HO}+\sum_{q\in\mathcal{Q}_-^g}\delta^{q}+\delta^g.
\end{eqnarray} 
Therefore, combining~\eqref{eq:404} and~\eqref{eq:040}, we conclude that $\hat{\phi_1}\geq \phi_1^*$, which contradicts our assumption. 

Thus in summary, the only possibility is that the effective total flow on lane 1 remains the same, i.e., $\hat{\phi}_1=\phi_1^*$, and the lane delays remain the same correspondingly.
\end{proof}

\begin{remark} [Interpretation of Theorem~\ref{thm:lane_delay_robust}]
    This theorem illustrates that lane delays can remain unaffected by vehicle misbehavior, provided the proportion of misbehaving vehicles remains sufficiently small. This resilience stems from the balance achieved when honest vehicles, detecting increased congestion in toll lane 1 due to misbehaving vehicles, opt for lane 2. Consequently, this adjustment by honest vehicles, though out of self-interest maximization, effectively absorbs the disruption caused by misbehaving vehicles, demonstrating a counter-intuitive capacity of the system to withstand a certain degree of misbehavior without deteriorating lane delays. In other words, by aiming to minimize their own costs, honest vehicles act as a buffer against the disruptions caused by misbehaving vehicles, mitigating the impact of toll violations on the overall system.
\end{remark}

Despite the conditions outlined in Theorem~\ref{thm:lane_delay_robust} ensuring that lane delays remain stable under moderate levels of vehicle misbehavior, the total commuter delay may still experience variations. This phenomenon is further elucidated in the subsequent theorem.
\begin{theorem} [The total commuter delay varies with vehicle misbehavior]
\label{theorem-JfhgJf}
Consider a tolled segment of highway $G=\left(\mathbf{D},\mathbf{d},n^{\text{LO}}, n^{\text{HO}},\mu,\bm \tau,\boldsymbol{\alpha}_{\rm m}\right)$ with differentiated tolls and potential vehicle misbehavior. Assume that conditions 1) and 2) in Theorem~\ref{thm:lane_delay_robust} hold. 

\begin{description}
\item[\textbf{a)}]  \label{th4-a} If $\mathcal{Q}^{g}_{+}\cap \mathcal{M}= \emptyset$, where  $\mathcal{Q}^{g}_{+}\subset \Bar{\mathcal{P}}:=\{p\in \Bar{\mathcal{P}}:\tau^p>\tau^g \} $ is the set that contains all vehicle classes with an assigned toll higher than $\tau^g$ and $\mathcal{M}:=\{p\subseteq \Bar{\mathcal{P}}:\alpha_{\rm m}^p>0\}$ is the set that contains the vehicle classes in which misbehavior occurs, then we have
\begin{align}
    J(\hat{\mathbf{f}}) = J(\mathbf{f}),
\end{align}
i.e., the total commuter delay of the lane choice equilibrium $\hat{\mathbf{f}}$ described in Definition~\ref{def:wdp_more}, which results as a consequence of vehicle misbehavior, is longer than the total commuter delay of the lane choice equilibrium $\mathbf{f}$ described in Definition \ref{def:wdp_diff}, which occurs without any vehicle misbehavior.

   \item[\textbf{b)}]  \label{th4-b} If $\mathcal{Q}^{g}_{+}\cap \mathcal{M}\neq \emptyset$ and every vehicle class $s\in  \mathcal{Q}^{g}_{+}\cap \mathcal{M}$ has a mobility degree $\nu^s<\nu^g$, then
   \begin{align}
    J(\hat{\mathbf{f}}) > J(\mathbf{f}).
\end{align}

    \item[\textbf{c)}]  \label{th4-c} If $\mathcal{Q}^{g}_{+}\cap \mathcal{M}\neq \emptyset$ and every vehicle class $s\in  \mathcal{Q}^{g}_{+}\cap \mathcal{M}$ has a mobility degree $\nu^s>\nu^g$, then
   \begin{align}
    J(\hat{\mathbf{f}}) < J(\mathbf{f}).
\end{align}
\end{description}

\end{theorem}

\begin{proof}
By Theorem \ref{thm:lane_delay_robust} and its proof, the lane delays of a lane choice equilibrium $\hat{\mathbf{f}}$ (with vehicle misbehavior)
will remain the same as the lane delays of the lane choice equilibrium $\mathbf{f}$ (without vehicle misbehavior).
Specifically, for any lane $i \in \mathcal{I}=\{1,2\}$, we have
\begin{eqnarray*}
\phi_i^* = \hat{\phi}_i \:\:\mbox{and}\:\:D_i\left(\phi_i^*\right) = D_i\left(\hat{\phi}_i\right),
\end{eqnarray*}
where $\hat{\phi}_i$ is the effective vehicle flow on lane $i$ at the lane choice equilibrium $\hat{\mathbf{f}}$ (with vehicle misbehavior), and $\phi_i^*$ is the effective vehicle flow at the lane choice equilibrium $\mathbf{f}$ (without vehicle misbehavior) specified in condition 1) in Theorem~\ref{thm:lane_delay_robust}.  Therefore, we have
\begin{align}
D_1\left(\hat{\phi}_1\right)+\tau^g=D_2\left(\hat{\phi}_2\right),
\end{align}
which implies that vehicles of class $g$ may travel on both lanes, while vehicles that are charged a toll lower than $\tau^g$ will choose to travel on lane 1, and honest vehicles that are charged with a toll higher than $\tau^g$ will choose to travel on lane 2.

Let $n^g$ be the commuter occupancy of class $g$ vehicles, i.e.,
\begin{align}
     n^g &= n^{\text{HO}},\ \text{if}\ g = \text{HV,HO},\\
     n^g &= n^{\text{LO}},\ \text{if}\ g \in \{\text{HV,LO; AV,LO}\}.
\end{align}
Given Eqs.~\eqref{eq:social_cost} and \eqref{eq:social_cost_more}, we have
\begin{align}
\label{eq:54}
    J\left(\mathbf{f}\right)-J\left(  \hat{\mathbf{f}}\right) 
    &= \left(\sum_{q\in \mathcal{Q}^{g}_{-}}d^q+n^gf_1^{g}+d^{\rm AV,HO}\right)D_1\left(\phi_1^*\right)+\left(\sum_{q\in \mathcal{Q}^{g}_{+}}d^q+n^gf_2^{g}\right)D_2\left(\sum_{p\in \mathcal{P}}\delta^p-\phi_1^*\right)\\
    &-\left[\left(\sum_{s\in \mathcal{M}\cap \mathcal{Q}^{g}_{+}}\alpha^{s}_{\rm m}d^{s}+\sum_{q\in \mathcal{Q}^{g}_{-}}d^q+\alpha^{g}_{\rm m}d^{g}+n^g\hat{f}_1^{g}+d^{\rm AV,HO}\right) D_1\left(\phi_1^*\right)\right.\nonumber\\
    &\ \ \ \ \ +\left.\left(\sum_{q\in \mathcal{Q}^{g}_{+}}d^q-\sum_{s\in \mathcal{M}\cap \mathcal{Q}^{g}_{+}}\alpha^{s}_{\rm m}d^{s}+n^g\hat{f}_2^{g}\right)D_2\left(\sum_{p\in \mathcal{P}}\delta^p-\phi_1^*\right)\right],\nonumber
\end{align}
where readers are reminded that the set $\mathcal{M}$ contains all misbehaving vehicle classes, i.e., for any vehicle class $s\in \mathcal{M}$, $\alpha^{s}_{\rm m}>0$. Rearranging terms in \eqref{eq:54}, we have
 \begin{align}\label{eq:555}
J\left(\mathbf{f}\right)-J\left(  \hat{\mathbf{f}}\right) 
     &= \left( n^gf_1^{g}-\sum_{s\in \mathcal{M}\cap \mathcal{Q}^g_{+}}\alpha^{s}_{\rm m}d^{s} -\alpha^{g}_{\rm m}d^{g}-n^g \hat{f}_1^{g} \right)D_1\left(\phi_1^*\right)\\
     &\ \ +\left( n^gf_2^{g}+\sum_{s\in \mathcal{M}\cap \mathcal{Q}^g_{+}}\alpha^{s}_{\rm m}d^{s} - n^g\hat{f}_2^{g} \right)D_2\left(\sum_{p\in \mathcal{P}}\delta^p-\phi_1^*\right).\nonumber
    \end{align}

Flow conservation implies that
    \begin{align}
        \hat{f}_1^{g}+\alpha^{g}_{\rm m}\delta^{g}+\hat{f}_2^{g}=f_1^{g}+f_2^{g},\ &\text{if}\ g \in \{\text{HV,LO; HV,HO}\},\label{eq:eqkey1}\\
        \mu\hat{f}_1^{g}+\alpha^{g}_{\rm m}\delta^{g}+\mu\hat{f}_2^{g}=\mu f_1^{g}+\mu f_2^{g},\ &\text{if}\ g=\rm AV,LO.\label{eq:eqkey2}
    \end{align}
Dependent on the specific class $g$, multiplying $\nu^g$, the mobility degree (Definition~\ref{def:md}) for vehicle class $g$, on both sides of either Eq.~\eqref{eq:eqkey1} or~\eqref{eq:eqkey2}, we have 
\begin{align}\label{eq:666}
    n^g\hat{f}_1^{g}+\alpha^{g}_{\rm m}d^{g}-n^gf_1^{g}=n^gf_2^{g}-n^g\hat{f}_2^{g}.
\end{align}

Moreover, since $\phi_1^*=\hat{\phi}_1$, we have 
\begin{align}
\label{eq:f1gmf1gh}
    f_1^{g}-\alpha^{g}_{\rm m}\delta^{g}-\hat{f}_1^{g}=\sum_{s\in \mathcal{M}\cap \mathcal{Q}^g_{+}}\alpha^{s}_{\rm m}\delta^{s}, \ &\text{if}\ g \in \{\text{HV,LO; HV,HO\}},\\
    \mu f_1^{g}-\alpha^{g}_{\rm m}\delta^{g}-\mu \hat{f}_1^{g}=\sum_{s\in \mathcal{M}\cap \mathcal{Q}^g_{+}}\alpha^{s}_{\rm m}\delta^{s}, \ &\text{if}\ g = \text{AV,LO}.
    \label{eq:f1gmf2gh}
\end{align}
Dependent on the specific class $g$, multiplying $\nu^g$ on both sides of Eq.~\eqref{eq:f1gmf1gh} or~\eqref{eq:f1gmf2gh}, we obtain 
\begin{align}\label{eq:777}
    n^g\hat{f}_1^{g}+\alpha^{g}_{\rm m}d^{g}-n^gf_1^{g}=\nu^g\sum_{s\in \mathcal{M}\cap \mathcal{Q}^g_{+}}\alpha^{s}_{\rm m}\delta^{s}.
\end{align}

Combining Eqs.~\eqref{eq:555} and~\eqref{eq:666}, we have
    \begin{align}
    \label{eq:55}
    J\left(\mathbf{f}\right)-J\left(  \hat{\mathbf{f}}\right)
    =&\left( n^g f_1^{g}-\sum_{s\in \mathcal{M}\cap \mathcal{Q}^{g}_{+}}\alpha^{s}_{\rm m}d^{s} -\alpha^{g}_{\rm m}d^{g}- n^g\hat{f}_1^{g} \right)D_1\left(\phi_1^*\right)\nonumber\\
    &+\left(  n^g\hat{f}_1^{g}+\alpha^{g}_{\rm m}d^{g}+\sum_{s\in \mathcal{M}\cap \mathcal{Q}^{g}_{+}}\alpha^{s}_{\rm m}d^{s}-n^gf_1^{g} \right)D_2\left(\sum_{p\in \mathcal{P}}\delta^p-\phi_1^*\right)\nonumber\\
    =&\left( n^g f_1^{g}-\sum_{s\in \mathcal{M}\cap \mathcal{Q}^{g}_{+}}\alpha^{s}_{\rm m}d^{s}- \alpha^{g}_{\rm m}d^{g}- n^g\hat{f}_1^{g} \right)\left( D_1\left(\phi_1^*\right)-D_2\left(\sum_{p\in \mathcal{P}}\delta^p-\phi_1^*\right)\right).
\end{align}
With Eq.~\eqref{eq:777}, we have
\begin{align}\label{eq:58}
    J\left(\mathbf{f}\right)-J\left(  \hat{\mathbf{f}}\right)
    =\left( \nu^g\sum_{s\in \mathcal{M}\cap \mathcal{Q}^g_{+}}\alpha^{s}_{\rm m}\delta^{s}-\sum_{s\in \mathcal{M}\cap \mathcal{Q}^{g}_{+}}\alpha^{s}_{\rm m}d^{s} \right)\left( D_1\left(\phi_1^*\right)-D_2\left(\sum_{p\in \mathcal{P}}\delta^p-\phi_1^*\right)\right).
\end{align}
According to condition 1) and Eq.~\eqref{eq:mis1} in Theorem \ref{thm:lane_delay_robust}, \begin{eqnarray}
\label{eq:D1D2}
 D_1\left(\phi_1^*\right)-D_2\left(\sum_{p\in \mathcal{P}}\delta^p-\phi_1^*\right) < 0.
\end{eqnarray}
Thus, as described in Theorem \ref{theorem-JfhgJf}, if $\mathcal{M}\cap \mathcal{Q}^{g}_{+}=\emptyset$,  we have
$J\left(\mathbf{f}\right)-J\left(  \hat{\mathbf{f}}\right)
    =0$. Otherwise if $\mathcal{M}\cap \mathcal{Q}^{g}_{+} \ne \emptyset$, for every vehicle class $s\in \mathcal{M}\cap \mathcal{Q}^g_{+}$, we have  $d^s = \nu^s\delta^s$ and
 \begin{align}\label{eq:59}
    J\left(\mathbf{f}\right)-J\left(  \hat{\mathbf{f}}\right)
    =\left(( \nu^g - \nu^s) \sum_{s\in \mathcal{M}\cap \mathcal{Q}^{g}_{+}}\alpha^{s}_{\rm m}\delta^{s}  \right)\left( D_1\left(\phi_1^*\right)-D_2\left(\sum_{p\in \mathcal{P}}\delta^p-\phi_1^*\right)\right) .
\end{align}
Additionally, if $\nu^s<\nu^g$,
by Eqs.~\eqref{eq:59} and~\eqref{eq:D1D2}, we have $J\left(\mathbf{f}\right)-J\left(  \hat{\mathbf{f}}\right)<0$. 
Conversely, if  $\nu^s>\nu^g$,  we have   $J\left(\mathbf{f}\right)-J\left(  \hat{\mathbf{f}}\right)>0$.
\end{proof}

\begin{remark} [Insights from Theorem~\ref{theorem-JfhgJf}]
Theorem~\ref{theorem-JfhgJf} provides two critical insights into how vehicle misbehavior influences the total commuter delay under different cases:
\begin{itemize}
    \item Case a) underscores that total commuter delay remains unaffected by vehicle misbehavior of vehicles assigned a toll smaller than or equal to $\tau^g$. 
    
    \item The outcomes observed in cases b) and c) are intricately linked to the specifics of the tolling policy implemented. Notably, in case c), the presence of vehicle misbehavior paradoxically leads to a decrease in total commuter delay. This suggests that the initial tolling strategy was not optimally configured, hinting at an irrational tolling approach. To rectify this, a more strategic tolling policy should be adopted, one that prioritizes vehicles with a larger mobility degree for toll lane 1 access by assigning them a lower toll. Such strategic tolling policies are showcased in case b) and detailed in Section~\ref{sec:diff_toll}.
\end{itemize}
\end{remark}

Theorems~\ref{thm:lane_delay_robust} and~\ref{theorem-JfhgJf} elucidate the influence of vehicle misbehavior on both lane delays and total commuter delay, respectively. To clarify these theoretical insights, we present a precise scenario demonstrating the outcomes of applying differentiated tolling in the presence of vehicle misbehavior.

\begin{figure*}[bht]
\centering
     \begin{subfigure}[b]{0.46\textwidth}
          \centering
         \includegraphics[clip, trim=0cm 0.7cm 0cm 0.5cm, width=1.00\textwidth]{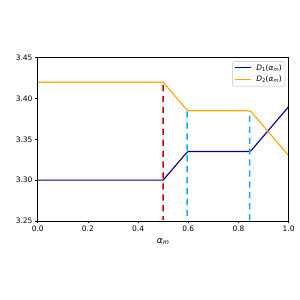}
         \caption{Lane delays: $D_1$-lane 1 (blue) and $D_2$-lane 2 (yellow)}
   \label{fig:counter_eg_a}
     \end{subfigure}
     \hfill
    \begin{subfigure}[b]{0.48\textwidth}
         \centering
         \includegraphics[clip, trim=0cm 0.8cm 0cm 0.5cm, width=\textwidth]{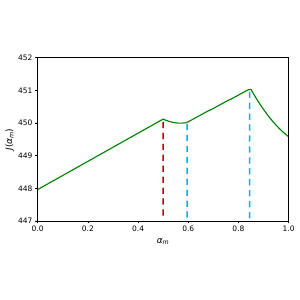}
         \caption{Total commuter delay}
     \label{fig:counter_eg_b}
    \end{subfigure}
    \caption{The lane delays and the total commuter delay are shown as functions of the proportion of misbehaving HV,LOs, $\alpha_{\rm m}^{\rm HV,LO}$, as described in Example~\ref{eg:mis}. The resilient regions are characterized by Theorem~\ref{thm:lane_delay_robust},~\ref{theorem-JfhgJf} and Proposition~\ref{prop:lane_delay_robust}.}
    \label{fig:counter_eg}
\end{figure*}

\begin{example}\label{eg:mis}
Consider a tolled highway segment with $\{d^{\rm AV,HO}=20,\ d^{\rm AV,LO}=30,\ d^{\rm HV,HO}=48,\ d^{\rm HV,LO}=36,\ n^\text{HO}=2,\ n^\text{LO}=1,\ \mu =0.3\}$ with the delay functions being the BPR functions in Eq. \eqref{eq:BPR}  with parameters $\mathbf{D}=\{\theta_i=3,\ \gamma_i=1,\ \beta_i=1,\ m_i=100:\ i\in \mathcal{I}\}$. Therefore, AV,HOs have the largest mobility degree given by $\nu^{\rm AV,HO}=\frac{n^\text{HO}}{\mu} \approx 6$. These vehicles travel freely on the toll lane 1. The other three classes of vehicles will be tolled while traveling on lane 1. AV,LOs have the largest mobility degree among the three classes of decision-making vehicles given by $\nu^{\rm AV,LO}=\frac{n^\text{LO}}{\mu} \approx 3$,  followed by the class of HV,HOs with a mobility degree $\nu^{\rm HV,HO}=n^\text{HO}=2$. HV,LOs have the smallest mobility degree $\nu^{\rm HV,LO}=n^\text{LO}=1$. A rational differentiated tolling policy should assign progressively lower tolls to vehicles as their mobility degrees increase. Thus, in this example, we assign the following tolls:  $\tau^{\rm HV,LO} = 0.3$, $\tau^{\rm HV,HO} = 0.12$ and $\tau^{\rm AV,LO} = 0.05$.
In this example, we consider the case when only HV,LOs misbehave i.e., $\alpha_{\rm m}^{\rm HV,LO}>0$, while $\alpha_{\rm m}^{\rm HV,HO}=\alpha_{\rm m}^{\rm AV,LO}=0$. 
 
From Eq.~\eqref{eq:mis1} in Condition 1) of Theorem~\ref{thm:lane_delay_robust}, we can determine that HV,HOs will be traveling on both lanes at the lane choice equilibrium, when no vehicles misbehave, i.e. $g = {\rm HV,HO}$. Correspondingly, according to the toll configuration, we have $\mathcal{Q}_-^g=\{\rm AV,LO\}$ and $\mathcal{Q}_+^g=\{\rm HV,LO\}$. 
We then gradually increase the proportion of misbehaving HV,LOs and plot the tendency curve of lane delays and the total commuter delay in Figure~\ref{fig:counter_eg}. 

Fig. \ref{fig:counter_eg_a}   plots lane delays as the proportion of misbehaving HV,LOs, $\alpha_{\rm m}^{\rm HV,LO}$, increases from $0$ to $1$.  As expected from the results of Theorem \ref{thm:lane_delay_robust},  lane delays remain robust to moderate levels of vehicle misbehavior until the proportion of misbehaving vehicles exceeds the bound given by Eq.~\eqref{eq:th3-alpha_bound}, which in this example is $\alpha_{\rm m}^{\rm HV,LO}=0.5$. When the bound~\eqref{eq:th3-alpha_bound} is exceeded, lane delays start to vary. Fig. \ref{fig:counter_eg_b}  plots the total commuter delay $J(\hat{\mathbf{f}})$ in Eq.~\eqref{eq:social_cost_more} as $\alpha_{\rm m}^{\rm HV,LO}$, increases from $0$ to $1$. Notice that condition b) in Theorem~\ref{theorem-JfhgJf} is satisfied. Aligned with the results of this theorem, the total commuter delay increases as $\alpha_{\rm m}^{\rm HV,LO}$ increases, until the bound ~\eqref{eq:th3-alpha_bound} is exceeded when $\alpha_{\rm m}^{\rm HV,LO}=0.5$ and the total commuter delay $J(\hat{\mathbf{f}})$ is no longer guaranteed  to be a monotonically increasing function of  $\alpha_{\rm m}^{\rm HV,LO}$.

\end{example}

Figs. \ref{fig:counter_eg_a} and \ref{fig:counter_eg_b} reveal that when the proportion of misbehaving vehicles, represented by $\alpha_{\rm m}^{\rm HV,LO}$ in this example, exceeds the limit specified in Eq.~\eqref{eq:th3-alpha_bound}, lane delays lose their previous resilience. This transition occurs because Equation~\eqref{eq:mis1} is breached when the threshold for misbehaving vehicles is exceeded. However, another phase of resilience emerges, exemplified by $\alpha_{\rm m}^{\rm HV,LO}\in [0.6,0.85]$ as illustrated in Fig. \ref{fig:counter_eg_a}, where lane delays once again become resilient. This return to resilience is contingent upon Equation~\eqref{eq:mis1} being fulfilled by a different vehicle class $g$.
These insights are comprehensively elaborated in the subsequent proposition.

\begin{proposition} [Characterization of more resilient regions of lane delays]\label{prop:lane_delay_robust}
Consider a tolled segment of highway $G=\left(\mathbf{D},\mathbf{d},n^{\text{LO}}, n^{\text{HO}},\mu,\bm \tau,\boldsymbol{\alpha}_{\rm m}\right)$ with differentiated tolls and potential vehicle misbehavior in Theorem~\ref{thm:lane_delay_robust}.
If there exists a vehicle class $g_-\in \mathcal{Q}^g_-$, i.e., $\tau^{g_-}<\tau^{g}$, then the lane delays remain resilient to vehicle misbehavior when the misbehaving vehicle proportions $\boldsymbol{\alpha}_{\rm m}$ lie in the following region:
    \begin{align}
    \label{eq:prop3-alpha_bound}
        \Tilde{\phi}_1 - \delta^{\rm AV,HO}- \sum_{q\in \mathcal{Q}^{g_-}_{-}}\delta^{q}-\delta^{g_-}(1-\alpha_{\rm m}^{g_-})\leq \sum_{p\in \Bar{\mathcal{P}}\setminus \mathcal{Q}^{g_-}_{-}}\delta^{p}\alpha_{\rm m}^{p}\leq \Tilde{\phi}_1 - \delta^{\rm AV,HO}- \sum_{q\in \mathcal{Q}^{g_-}_{-}}\delta^{q},
    \end{align}
   where the effective vehicle flow $\Tilde{\phi}_1$ is determined by the following equation:
   \begin{align}\label{eq:prop3_phi}
       D_1\left(\Tilde{\phi}_1\right)+\tau^{g_-}=  D_2\left(\sum_{p\in \mathcal{P}}\delta^{p}-\Tilde{\phi}_1\right).
    \end{align}

Furthermore, consider the total commuter delay $J$ as a function of $\Tilde{\alpha}:=\sum_{p\in  \mathcal{Q}^{g_-}_{+}}\delta^{p}\alpha_{\rm m}^{p}$. We have the following conclusions.
    \begin{description}
\item[\textbf{a)}]  \label{prop3-a} If $\mathcal{Q}^{g_-}_{+}\cap \mathcal{M}= \emptyset$, then $J(\Tilde{\alpha})$ remains constant within the region characterized by Eq.~\eqref{eq:prop3-alpha_bound}.
   
   \item[\textbf{b)}]  \label{prop3-b} If $\mathcal{Q}^{g_-}_{+}\cap \mathcal{M}\neq \emptyset$, and every vehicle class $s\in  \mathcal{Q}^{g_-}_{+}\cap \mathcal{M}$ has a mobility degree $\nu^s<\nu^{g_-}$, then $J(\Tilde{\alpha})$ is an increasing function within the region characterized by Eq.~\eqref{eq:prop3-alpha_bound}.

    \item[\textbf{c)}]  \label{prop3-c} If $\mathcal{Q}^{g_-}_{+}\cap \mathcal{M}\neq \emptyset$, and every vehicle class $s\in  \mathcal{Q}^{g_-}_{+}\cap \mathcal{M}$ has a mobility degree $\nu^s>\nu^{g_-}$, then $J(\Tilde{\alpha})$ is a decreasing function  within the region characterized by Eq.~\eqref{eq:prop3-alpha_bound}.

\end{description}

\end{proposition}

\begin{proof}
The proof of the proposition can be obtained by replacing vehicle class $g$ with vehicle class $g_-$ within the proof of Theorem~\ref{thm:lane_delay_robust} and~\ref{theorem-JfhgJf}. Thus the proof is omitted here.
\end{proof}

Referring back to Example~\ref{eg:mis} and guided by Proposition~\ref{prop:lane_delay_robust}, we note that with $\mathcal{Q}_-^g=\{\rm AV,LO\}$ indicating a single vehicle class $g_-=\rm AV,LO$, only one additional region of resilience is anticipated beyond the resilient region $\alpha_{\rm m}^{\rm HV,LO} \leq 0.5$ specified in Theorem~\ref{thm:lane_delay_robust}. The additional region of resilience can be characterized by first calculating  $\Tilde{\phi}_1 =33.5$, using Eq.~\eqref{eq:prop3_phi}, and then by identifying the region's bounds, $\alpha_{\rm m}^{\rm HV,LO}\in [0.6,0.85]$, per Eq.~\eqref{eq:prop3-alpha_bound}. Within this resilient zone, lane delays remain constant, as illustrated in Fig.~\ref{fig:counter_eg_a}. Furthermore, aligning with Condition b) of Proposition~\ref{prop:lane_delay_robust}, we expect that the total commuter delay exhibits a monotonic increase as $\alpha_{\rm m}^{\rm HV,LO}$ increases, as depicted in Fig.~\ref{fig:counter_eg_b}.

Our analysis in this section, encompassing both qualitative and quantitative aspects, focuses exclusively on identified resilience regions. The phenomena occurring outside these regions, such as within the intervals $\alpha_{\rm m}^{\rm HV,LO}\in (0.5, 0.6)$ and
$\alpha_{\rm m}^{\rm HV,LO}\in (0.85,1]$ in Fig.~\ref{fig:counter_eg_b}, are not covered in this study.  
However, in this simplified example, the observed tendency of the delays can be elucidated as follows. With an increase in toll violations by misbehaving vehicles, legitimate users of lane 1 opt for lane 2, to minimize their cost. The transition points of vehicle misbehavior, $\alpha_{\rm m}^{\rm HV,LO} = 0.5$ and $\alpha_{\rm m}^{\rm HV,LO} = 0.85$, mark the instances where the total commuter delay reaches a local maximum as a consequence of all honest vehicles from a specific class being driven out of lane 1. Specifically, in this example, all vehicles of class HV,HO are driven out of lane 1 when $\alpha_{\rm m}^{\rm HV,LO} = 0.5$, while all vehicles of class AV,LO are driven out of lane 1 when $\alpha_{\rm m}^{\rm HV,LO} = 0.85$. After this occurs, honest vehicles' buffering ability is suspended, therefore, misbehaving vehicles' impact outweighs and causes the downward slope of total commuter delay as a function of vehicle misbehavior. The period ends when another class of honest vehicles reaches a mixed strategy equilibrium and starts buffering again.
It's important to underscore that while during the period, increased misbehavior momentarily ameliorates total commuter delay, it invariably results in a higher total commuter delay compared to a scenario with no misbehavior, alongside revenue losses for toll-collection agencies.

\section{Conclusions and Future Work} \label{sec:future}
In this study, we introduced a toll lane framework for a highway segment shared by four distinct vehicle classes, including a restricted/tolled lane. Autonomous vehicles with high occupancy, possessing the highest mobility degree and thus significantly enhancing commuter mobility, are granted free access to this lane. Meanwhile, the other three vehicle classes—human-driven high-occupancy, autonomous low-occupancy, and human-driven low-occupancy vehicles—face a choice: pay a toll to use the restricted lane or opt for the toll-free regular lanes. Operating under the assumption that vehicles act selfishly, prioritizing cost minimization, we analyzed the resulting lane choice equilibria. Our focus was on comparing the capabilities of high-occupancy versus autonomous vehicles in enhancing social mobility and formulating efficient rational tolling strategies. Our framework offers a comprehensive approach for designing and evaluating toll lane policies, encompassing the determination of an optimal uniform toll, defining the occupancy threshold for high-occupancy vehicle classification, and outlining efficient differentiated tolling strategies to minimize overall commuter delay. Additionally, we examined the impact of vehicles' toll violation behavior and revealed the framework's inherent resilience.

This study illuminates the broader implications of human-autonomy interactions and their impact on traffic management:

\begin{itemize}
    \item Dual Nature of Interaction: Contrary to the common narrative where the selfish behavior of human drivers typically detracts from overall system efficiency, our findings reveal a paradoxical role where such behavior inadvertently acts as a buffer, mitigating potential disruptions. This observation opens up avenues for further exploration into the multifaceted effects of human-autonomy interaction across diverse scenarios and objectives.
    \item 
Revolutionizing Infrastructure for Autonomy Integration: Our framework offers significant contributions towards evolving current infrastructure to accommodate the seamless integration of autonomous vehicles. By providing key insights and aligning with a broad spectrum of related considerations, it lays the groundwork for future endeavors in infrastructure transformation, ensuring compatibility with emerging autonomous technologies.
\end{itemize}

These insights underscore the complexity of integrating autonomous vehicles into existing traffic systems and highlight the need for comprehensive strategies that account for both the challenges and opportunities presented by this transition.

\section*{Acknowledgments}
This work was supported by the National Science Foundation under Grants CPS 1545116 and ECCS-2013779.


\bibliographystyle{IEEEtran}
\bibliography{MLU.bib,onramp.bib,CDC2020.bib,protocol.bib,cdc.bib}

\end{document}